\documentclass[12pt]{article}

\usepackage[utf8]{inputenc}
\usepackage[T1]{fontenc}
\usepackage{amsmath,amssymb,amsthm}
\usepackage{graphicx}
\usepackage[hidelinks]{hyperref}
\usepackage[numbers,sort&compress]{natbib}
\usepackage[margin=1in]{geometry}
\usepackage{newtxtext,newtxmath}
\usepackage{setspace}
\usepackage{float}
\usepackage{caption}
\usepackage{ragged2e}
\usepackage{upgreek}
\usepackage{xurl}

\newcommand{\um}{\,µm}
\newcommand{\nm}{\,nm}

\newcommand{\keywords}[1]{%
    \noindent\textbf{Keywords: }#1\par
}

\makeatletter
\renewcommand{\fnum@figure}{\textbf{Figure \thefigure}}
\makeatother

\justifying
\date{}

\begin{document}

\begin{center}

{\Large\bfseries High-Resolution X-ray Spectro-Microscopy of Au Microcrystallites\par}
\vspace{1em}

Chaitali Sow\textsuperscript{1*},
Martin Seyrich\textsuperscript{1},
Johannes Hagemann\textsuperscript{1},
Giridhar U. Kulkarni\textsuperscript{2},
Thomas F. Keller\textsuperscript{1,3},
Christian G. Schroer\textsuperscript{1,3,4},
Andreas Schropp\textsuperscript{1}

\vspace{1em}

{\small
\textsuperscript{1} Centre for X-ray and Nano Science CXNS, Deutsches Elektronen-Synchrotron DESY, Notkestr. 85, Hamburg 22607, Germany\par
\vspace{0.5em}
\textsuperscript{2} Chemistry and Physics Of Materials Unit (CPMU), Jawaharlal Nehru Centre For Advanced Scientific Research (JNCASR), Rachenahalli Lake Rd, Jakkur, Bengaluru, Karnataka 560064, India\par
\vspace{0.5em}
\textsuperscript{3} Center for Hybrid Nanostructures, Institute for Nanostructure and Solid-State Physics, Universität Hamburg, Luruper Chaussee 149, 22761 Hamburg, Germany\par
\vspace{0.5em}
\textsuperscript{4} Helmholtz Imaging, Deutsches Elektronen-Synchrotron DESY, Notkestr. 85, Hamburg 22607, Germany\par
}

\vspace{2em}

{\small
\textsuperscript{*} Email:
\href{mailto:chaitali.sow@desy.de}{chaitali.sow@desy.de}, 
\href{mailto:chaitali.sow@xray-lens.de}{chaitali.sow@xray-lens.de}
\par
}
\end{center}


\justifying
\vspace{1cm}
\noindent Understanding the functionalities of heterogeneous catalysts requires high-resolution nanoscale chemical imaging tools.
X-ray spectro-ptychography offers this capability by enabling X-ray absorption near edge structure (XANES) based amplitude and phase contrasts imaging with high spatial resolution.
However, such demonstrations at energies above \text{10 keV}, such as the Au L-edges, remain limited due to weak X-ray absorption cross-sections and correspondingly, low signal levels.
Here, multimodal spectro-microscopy has been performed on Au microcrystallites that host both conventional face-centered cubic (\textit{fcc}) and unconventional non-cubic lattices, particularly important for catalysis.
Amplitude and phase contrasts across the $\mathrm{L}_3$ edge are resolved on thin Au crystallites (150\nm{} to 1.71\um{}) with 45 nm spatial resolution.
The resulting XANES spectra exhibit excellent agreement with standard transmission and fluorescence contrasts, while providing complementary chemical insights with improved spatial resolution.
Quantitative retrieval of anomalous phase variations \mbox{$\sim$12 keV} further demonstrates the suitability of this method for high-resolution chemical imaging across a wide range of functional materials.

\section*{Introduction}

With advancements of nanoscience and nanotechnology, the requirement of high resolution characterization tools has steadily increased.
This need is particularly evident in materials science, including catalysis and battery research.
For example, catalysts composed of metal nanostructures, such as Ir for green hydrogen production\cite{lee2021high}, Pt for automotive catalytic converters\cite{omrani2020platinum} and petrochemical refining\cite{liu2017generation}, and Au for low temperature CO oxidation\cite{grunwaldt1999comparative} and $\mathrm{CO_2}$ reduction\cite{wang2020undercoordinated}, are widely used in energy and chemical technologies.
Over time, the nanostructures can undergo modification or degradation, which directly affects their overall performance\cite{zugic2017dynamic,bonavia2025pt}.
Therefore, characterization of nanostructures is prerequisite for better design of catalysts as well as understanding their behavior, including under in-situ and operando conditions.
Moreover, nanostructures often host strain, exhibit different lattices and display intrinsic non-homogeneity\cite{Sow2020Noble}.
As a result, investigation of their electronic structure and chemical information with high-spatial resolution is necessary for establishing detailed structure-property correlation.\\
Electron energy loss spectroscopy (EELS) in scanning transmission electron microscopy offers high-resolution chemical/electronic structure mapping.
For instance, monochromated EELS has been used
to reveal distinct optical properties in hexagonal close-packed (\textit{hcp}) Au nanocrystals\cite{Peng2019Plasmons} and nanorods\cite{fan2020heterophase}.
The variations in plasmon frequency of \textit{hcp} Au nanocrystals compared to its conventional \textit{fcc} lattice are attributed to differences in free electron density and plasmon decay pathways, reflecting their unique electronic band structures\cite{Peng2019Plasmons}.
Despite the high resolution information EELS provides, the technique is limited to relatively thin nanostructures (less than 100 nm)\cite{Peng2019Plasmons, meledina2023multimodal,loy2024advanced}.
Although, it is feasible to perform EELS studies with the help of a focused ion beam (FIB), the latter technique is destructive and can induce damage within the nanostructures, as observed in Au microcrystallites\cite{Sow2020Unraveling}.\\
X-ray absorption spectroscopy (XAS) is routinely used for understanding the chemical/electronic and structural composition of materials, although without any spatial information.
Its combination with microscopy \textit{i.e.}, spectro-microscopy yields local XAS contrast with high spatial resolution\cite{grunwaldt2010hard,Price2015Chemical,Matsui2022Variation}.
Currently employed spatially resolved X-ray based non-destructive imaging techniques are scanning X-ray diffraction microscopy (SXDM)\cite{thibault2008high,Sow2020Unraveling}, transmission X-ray microscopy (TXM)\cite{gao2024Transmission}, scanning transmission X-ray microscopy (STXM)\cite{de2008nanoscale,Price2015Chemical}, X-ray fluorescence (XRF)\cite{segura2011nano} and ptychography\cite{Dierolf2010Ptychographic,sasaki2025development}.
The spatial resolution of conventional scanning techniques, such as STXM, XRF, and SXDM, is limited by the probe size, commonly to $\sim$100 nm.
In contrast, in ptychography, the sample is scanned and a far-field diffraction pattern is acquired at each scan point.
By scanning with an overlap between the illuminations of neighboring points, the complex transmission function of the sample can be iteratively reconstructed\cite{Dierolf2010Ptychographic,Pfeiffer2018X-ray,guizar2021ptychography} with access to both phase and amplitude.
The spatial resolution of ptychography, a lensless imaging technique, is limited only by the maximum scattering angle at which a statistically relevant signal in the diffraction pattern is measured\cite{schropp2010hard}.
Consequently, the maximum resolution achieved so far with ptychography, typically few tens of nanometers, significantly surpasses that of conventional X-ray imaging techniques, as it is not constrained by the X-ray optics employed.
Energy- dependent TXM/STXM and XRF across the absorption edge of an element allow to measure XAS with transmission\cite{ishiguro2023comprehensive,ishiguro2023oxidation,arul2021review} and fluorescence\cite{segura2011nano,Price2015Chemical,Pattammattel2020High} contrasts, respectively.
On the other hand, due to anomalous scattering, from the spectro-ptychography data set (\textit{i.e.}, combining X-ray ptychography with XAS), energy-dependent change in phase and amplitude (\textit{i.e.}, transmission) across the absorption edge can be extracted\cite{Urquhart2022X-ray}.\\
Spatially resolved chemical imaging  of nanostructures with both phase and absorption contrasts have been successfully demonstrated by soft and tender X-ray ptychography across research fields including chemistry\cite{maiden2013soft,Wise2016Nanoscale,Wu2018High,Sun2021Soft}, materials science\cite{Shapiro2014Chemical,shapiro2020ultrahigh}, biology\cite{Beckers2011Chemical, sakurai2025chemical} as well as in geoscientific and environmental research\cite{Lo2019Multimodal,Forster2021X-ray}.
In the materials community, hard X-ray spectro-ptychography is recently being used for the characterization of catalysts\cite{Hirose2018Visualization,Ihli2018Resonant,Gao2021Sparse,quinn2024optimal}, battery materials\cite{uematsu2021visualization,ishiguro2023comprehensive, Uematsu2023Nanoscale,sasaki2025development} and magnetic specimens\cite{donnelly2016high,donnelly2017three}.
For example, the Ce valance distribution in a Pt-supported cerium-zirconium oxide catalyst has been mapped using amplitude contrast spectro-ptychography\cite{Hirose2018Visualization}. Visualization of three-dimensional heterogeneities within an industrially used vanadium phosphorus oxide catalyst has been demonstrated with phase-contrast spectro-tomography\cite{Gao2021Sparse}.\\
Although the phase and absorption contrasts are generated simultaneously in ptychography, the phase contrast is typically stronger at the hard X-ray ranges with better signal-to-noise ratio as compared to the absorption contrast, mainly due to the weak absorption in thin nanostructures around the absorption edge.
As a result, spectro-ptychography of thin nanostructures above \text{10 keV} remains limited, making it challenging to characterize materials composed of Au, Ir and Pt, whose $\mathrm{L}_3$ absorption edges that lie above \text{11 keV}.
For example, chemical imaging of Au nanoparticles has been demonstrated using phase-contrast spectro-ptychography at approximately 11.89 to 11.94 keV, although the spectrum was notably noisy\cite{Hoppe2013High,Reinhardt2018Hard}.
A meaningful spectrum of the amplitude was not successfully generated as a result of the low signal-to-noise ratio.
Consequently, quantitative spectro-ptychography imaging of thin nanostructures at energies above $\text{10}\ \text{keV}$ remains uncommon.\\
Herein, we study micrometer-thick \textit{fcc} Au crystallites and sub-micrometer-thick unconventional body-centered orthorhombic and tetragonal (together, \textit{bc(o,t)}) Au crystallites using spectro-ptychography across the Au $\mathrm{L}_3$ absorption edge.
The \textit{bc(o,t)} crystallites demonstrate superior catalytic performance in the model ${\textit{p}}$-nitrophenol reduction reaction\cite{Mettela2017Non}.
Additionally, the \textit{bc(o,t)} Au crystallites exhibit an order of magnitude lower dissolution in aqua regia, in contrast to the conventional \textit{fcc} Au \cite{Mettela2018Nobler, Sow2021Crystal}.
The pronounced structure-dependent properties observed prompt the question: how are the electronic structures of \textit{bc(o,t)} Au influenced by their crystal structures?
Hitherto, the electronic structures of the \textit{bc(o,t)} Au crystallites remain largely unknown.\\
Therefore, we perform multimodal imaging of the Au crystallites by combining spectro-ptychography with STXM, XRF, and SXDM.
Although XANES generally refers to absorption, in the present context, it is taken to represent transmission and phase contrasts as well.
The parallel measurement of different contrasts as complementary techniques with high spatial resolution improves statistics and consequently results in the validation of the findings with great confidence\cite{grunwaldt2013imaging,sullivan2025spectro}.
In this way, XANES spectral information can be concurrently extracted from ptychographic phase, ptychographic amplitude, X-ray transmission and X-ray fluorescence contrasts.
The study showcases the possibility of chemical imaging using these modalities, with spatial resolutions of $\text{45}\ \text{nm}$ (ptychography) and $\text{100}\ \text{nm}$ (STXM/XRF).

\section*{Results}
\subsection*{Multimodal XANES using XRF, STXM and ptychography}
The synthesized \textit{bc(o,t)} Au microcrystallites are majorly in bipyramid, tetrahexahedral (THH) and hexagram shapes\cite{Mettela2015Ambient,Sow2021Crystal,Puliyassery2021optical}.
The bipyramids are approximately 3\um{} in length and about 400\nm{} in width, whereas the THH and hexagram are approximately 500\nm{} wide (cf. Figure S1).
Recent SXDM investigation of a single \textit{bc(o,t)} bipyramid crystallite showcased the co-presence of \textit{fcc} and \textit{bc(o,t)} lattices\cite{Sow2020Unraveling, sow2025real}.
Particularly, the tips of the bipyramid are enriched with \textit{fcc} lattice while the body is enriched with \textit{bc(o,t)} lattices.
Similarly, selected area electron diffraction showcases the corners of the tetrahexahedral (THH) and hexagram shaped crystallites host the \textit{fcc} lattice\cite{Mettela2017Non}.
Therefore, all the shapes host both \textit{fcc} and \textit{bc(o,t)} lattices\cite{Sow2024Deciphering}.\\
In order to study Au microcrystallites hosting different morphologies and crystal structures, individually synthesized and characterized ${\textit{fcc}}$ and ${\textit{bc(o,t)}}$ crystallites as in \mbox{Figure S1} were drop casted on a Si/SiN\textsubscript{x} substrate.
The middle of the substrate (500 × 500\um\textsuperscript{2})\ hosts a 100\nm{} thick X-ray transparent SiN\textsubscript{x} membrane.
The microcrystallites were located by SEM (\mbox{Figure S2}) and then subjected to the XANES study with the X-ray microscope (PtyNAMi) at beamline P06, PETRA III, DESY\cite{Schropp2020PtyNAMi}.
The sample was positioned in the focal plane and scanned in a 2D overlapping grid with a step size of (80 × 40\nm\textsuperscript{2})\ along the y-z direction with a total acquisition time of \text{6 min} (Figure \ref{fig1}a).
The experimental setup enabled simultaneous collection of STXM, XRF, SXDM, and ptychography data (Figure \ref{fig1}a-e). 
STXM contrast was obtained by integrating the diffraction signal of the far-field diffraction patterns collected for ptychography.
The 2D scan was repeated for the total energy window of \text{11.842 to 11.882 keV} in \text{2 eV} steps and followed by 11.883 to 11.972 keV in \text{1 eV} steps spanning the Au $\mathrm{L}_3$ XANES region while keeping the scanning parameters (\textit{i.e.}, grid size, step size and acquisition time) the same.
Therefore, the experiment generates a 3D data set where the third dimension is the energy (\mbox{Figure \ref{fig1}b-e}).
A XANES spectrum can be derived from the 3D data set at each pixel of the fluorescence, transmission, phase-contrast and squared-amplitude-contrast maps (\mbox{Figure \ref{fig1}f,g}).
\\The studied sample consist of a 3D-shaped crystallite, a big-sized tetrahexahedral (THH), two small THHs (named hereafter as $\text{THH}_1$ and $\text{THH}_2$) and a thin bipyramid crystallite which are 2.4, 1.4, 0.5, 0.5 and  2.2\um{} in length, respectively (see \mbox{Figure \ref{fig2}a}).
The ${\textit{fcc}}$ crystallites are larger whereas the ${\textit{bc(o,t)}}$ crystallites are smaller in size (also see \mbox{Figure S1b-e})\cite{Mettela2013Highly, Mettela2015Ambient, Mettela2017Non} and therefore, the big-sized THH and 3D shaped crystallite as shown in \mbox{Figure \ref{fig2}a} host an ${\textit{fcc}}$ lattice while the smaller THHs and the bipyramid are in ${\textit{bc(o,t)}}$ rich lattices.\\
The XRF and STXM maps at ${\text{11.963\ keV}}$ are shown in \mbox{Figure \ref{fig2}b,c}.
The stronger fluorescence intensities from the bigger size crystallites relate to their higher thicknesses.
Similarly, the weaker transmission intensities from $\mathrm{THH}_1$, $\mathrm{THH}_2$ and the bipyramid indicate that they are thin (\mbox{Figure \ref{fig2}c}).
The fluorescence spectrum is fitted using PyMCA\cite{Sole2007Multiplatform} and the fluorescence contribution from the Au $\mathrm{L}_{\upalpha}$ line is taken into consideration.
In the case of transmission, the STXM map is normalized with respect to the background (black dashed box in \mbox{Figure \ref{fig2}c}).
Ptychographic reconstruction of the complex-valued sample transmission function is performed from far-field diffraction patterns using ePIE\cite{Maiden2009Improved}.
The complex refractive index \textit{n} can be expressed as:
\begin{equation}
    n = 1 - \delta+i\beta,
    \label{eq1}
\end{equation}

where the real ($1-{\delta}$) and imaginary (${\beta}$) parts describe refraction and absorption, respectively.
Ptychographic reconstructions have a global phase ambiguity and, therefore, the background was fitted and subtracted as shown in \mbox{Figure S3} (further details in experimental section).
The lowest measurable phase shift with the PtyNAMi setup is about \( \pm 5\ \text{mrad}\)\cite{Reinhardt2018Hard}.
From the phase map (at 11.963 keV) as shown in \mbox{Figure \ref{fig2}d}, the ${\textit{fcc}}$ 3D-shaped crystallite and ${\textit{fcc}}$ THH show a maximum phase shift of ${\text{-1.937}}$ and ${\text{-1.474\ rad}}$, respectively.
Besides, the phase shift values for ${\textit{bc(o,t)}}$ $\text{THH}_1$ and $\text{THH}_2$ are relatively higher than the bipyramid; ${\text{-0.476}}$, ${\text{-0.471}}$ and ${\text{-0.375\ rad}}$, respectively.
The tip region of the bipyramid accounts for the smallest phase shift of only ${\text{-0.068\ rad}}$.
Similar to the STXM, the squared of the amplitude, \textit{i.e.}, transmission map from ptychographic reconstruction is normalized with respect to the background and is shown in \mbox{Figure \ref{fig2}e}.
Therefore, both directly measured and reconstructed transmission maps are available (\mbox{Figure \ref{fig2}c,e}).
Spatial resolution of the XRF and STXM is $\sim$100 nm with the pixel size of $\sim$40 nm (limited by the step size).
On the other hand, the resolution in the ptychographic reconstruction is ${\text{45\ nm}}$ (see \mbox{Figure S4}) with the pixel size of ${\text{17\ nm}}$.
Evidently, phase-contrast imaging is more sensitive towards small and thin samples when compared with amplitude contrast at the high energies (see Figure \ref{fig2}d,e).\\
The phase shift map away from the absorption edge, at ${\text {11.842\ keV}}$, as shown in \mbox{Figure \ref{fig3}a}, is considered for the calculation of the thickness of the sample (\mbox{Figure \ref{fig3}b}).
The thickness values are obtained following Equation (2)
\begin{equation}
    d=\frac{{|\phi|}{\lambda}}{2{\pi}{\delta}},
    \label{eq2}
\end{equation}

where ${\lambda}$, d and ${\phi}$ are the X-ray wavelength, thickness, and phase shift, respectively.
The index of refraction decrement ${\delta}$ of \textit{fcc} Au is obtained from the Henke database and is \text{1.9048 × 10\textsuperscript{-5}} at \text{11.842 keV}\cite{henke1993x,Xray_Interactions}.
The error bar of ${\phi}$ is \(\pm 5\ \text{mrad}\) (uncertainties of measurable phase), corresponding to a thickness uncertainty of ${\pm 4\ \text{nm}}$.
Line profiles drawn over differently shaped crystallites as in \mbox{Figure \ref{fig3}b} are shown in \mbox{Figure \ref{fig3}c,d}.
Particularly, ${\textit{bc(o,t)}}$ THHs are thicker than the ${\textit{bc(o,t)}}$ bipyramid while thinner than ${\textit{fcc}}$ THH.
The quantified thickness values are in agreement with the 2D projected SEM image in \mbox{Figure \ref{fig2}a}.
Therefore, this is a sample consisting of Au crystallites with different shapes, thicknesses, and crystal structures, an ideal sample system to study by spectro-microscopy.\\
To extract quantitative information, background subtracted\slash normalized maps are considered throughout for various contrasts.
Besides, the images over the energy stack are aligned for the extraction of spectral information with better spatial resolution, using rigid and B-spline transformations\cite{Bron2014Fast, Ntatsis2023itk}.
From the fluorescence, transmission (both STXM\slash ptychography) and phase stacks, integrated and averaged XANES spectra over each crystallite are extracted and plotted in \mbox{Figure \ref{fig4} and S5}.
For the 3D-shaped crystallite, the XANES spectra collected over the energy stack of fluorescence (XRF) and transmission (STXM) contrasts are shown in \mbox{Figure ${\ref{fig4}}$a}.
The energy-dependent phase shift for the same 3D-shaped crystallite is plotted in \mbox{Figure \ref{fig4}b}, which shows a well-formed peak with peak maximum (${\text{-1.126\ rad}}$) at \makebox{11.919 keV}.
Therefore, the anomalous phase change at the edge is ${\text{0.135}}$ rad, \textit{i.e.}, a relative change of ${\text{10.7}}{\%}$.
Along with phase shift, energy-dependent transmission (obtained from ptychographic reconstruction) is also plotted in \mbox{Figure \ref{fig4}b}. 
As expected, this spectrum resembles the transmission spectrum from STXM as shown in \mbox{Figure \ref{fig4}a}.
In order to correlate the experimentally obtained results from various contrasts, the normalized XANES spectra from fluorescence and transmission (STXM\slash ptychographic reconstruction) contrasts of the 3D-shaped crystallite are compared (see \mbox{Figure \ref{fig4}c}).
All the spectra nicely overlay and therefore, the generated XANES spectra from the various measured contrasts provide similar information.\\
Similarly, integrated and averaged XANES spectra are extracted from the small sized ${\textit{bc(o,t)}}$ $\text{THH}_2$ (see \mbox{Figure \ref{fig4}d,e}).
An absorption spectrum can be seen in fluorescence contrast while the transmission spectra from both STXM and ptychographic reconstruction clearly show the XANES features.
In contrast to the 3D-shaped crystallite, a smaller edge jump is visible in the transmission spectra of $\text{THH}_2$.
XANES from phase contrast is also derived in \mbox{Figure \ref{fig4}e} and the peak at 11.919 keV accounts for $\text{0.031 rad}$ change in phase shift.
Therefore, the measured data shows a distinct XANES spectrum even in the $\text{420 nm}$ thick $\text{THH}_2$ crystallite.
Particularly, the reconstructed transmission spectrum is more sensitive in showcasing the XANES features than the spectrum collected from STXM (compare \mbox{Figure \ref{fig4}d,e}).
Further, the spectral details pertained to ${\textit{fcc}}$ THH, ${\textit{bc(o,t)}}$ $\text{THH}_1$, and ${\textit{bc(o,t)}}$ bipyramid crystallites are shown in \mbox{Figure S5}.
Irrespective of the thickness of the crystallite (1.71\um{} to $\text{366 nm}$), the fluorescence spectra and their features are very similar (see \mbox{Figure \ref{fig4}a,d and S5a,c,e}).
Moreover, the changes in phase shift at the absorption edge are ${\text{0.087}}$, ${\text{0.028}}$ and $\text{0.026 rad}$ for ${\textit{fcc}}$ THH, ${\textit{bc(o,t)}}$ $\text{THH}_1$ and ${\textit{bc(o,t)}}$ bipyramid crystallites, respectively.
Given different crystallites have different thicknesses, the anomalous phase changes at the edge are ${\sim}$\text{8.2 to 10.8}\% for all the crystallites, where the lower phase shift is noticed for smaller crystallites.
Notably, the XANES with phase shift and amplitude contrasts are also well-resolvable even for the thinnest bipyramid (\mbox{Figure S5}).
The reduction in transmission intensity at the absorption edge (by STXM and ptychographic reconstruction) is seen for all the crystallites, although weak for the thinner crystallites.
The small changes in transmission intensity at the edge jump are well understood due to the lower absorption by the thinner crystallites.
The XANES (from energy-dependent fluorescence) spectra collected from differently shaped crystallites are compared in \mbox{Figure \ref{fig4}f}.
The spectra exhibit good agreement across all crystallites, indicating that the integrated and averaged XANES spectra are similar regardless of size and shape.\\
In order to identify the position of the absorption edge, \mbox{1$^{\text{st}}$} derivative of the XANES (fluorescence contrast) spectrum is considered along with the energy-dependent phase shift for 3D-shaped crystallite.
Particularly, the derivative of the absorption spectrum (fluorescence contrast) of the 3D-shaped crystallite highlights a peak at ${\text{11.919\ keV}}$ and, therefore, is the position of the edge.
Besides, the edge position obtained from the peak maximum of the phase shift is also at the same position (\mbox{Figure \ref{fig4}b}).\\
In small-angle scattering geometry, the interaction between X-rays and the electrons of the crystallite enables the complex refractive index, \textit{n}, to be represented using the atomic scattering factor in the forward direction $f(q \approx 0) = f_1 + i f_2$:

\begin{equation}
    n = 1-\frac{{r_e}{n_a}}{2\pi} \lambda^2(f_1+if_2), \label{eq3}
\end{equation}

where \textit{q}, $r_e$, $n_{a}$ and ${\lambda}$ are the momentum transfer vector, classical electron radius, atomic density, and wavelength, respectively (see Equation 1 and 3).
Therefore, $f_1$ and $f_2$ are proportional to refraction decrement $\delta$ and the attenuation increment $\beta$, respectively (cf. Equation 1).\\
The dispersion relation of phase and amplitude are related by the Kramers-Kronig relations (KKR) and one can be retrieved from the other\cite{Kronig1925Dispersion}.
Consequently, $f_1$ can be extracted from the ptychographic transmission signal (\textit{i.e.}, squared of the transmitted amplitude)\cite{Watts2014Calculation} and is plotted alongside the phase shift values in \mbox{Figure \ref{fig4}b}.
The derived $f_1$ spectrum closely follows the phase-shift spectrum, showing a good consistency between the transmission and anomalous dispersion.
Likewise, the absorption edge has been identified for ${\textit{bc(o,t)}}$ $\text{THH}_2$ by fluorescence and phase contrasts (\mbox{Figure \ref{fig4}e}) at \text{11.919 keV}.
The calculated $f_1$ is found to complement the phase-shift spectrum with the absorption edge at 11.919 keV for all other crystallites (see \mbox{Figure S5}).
Therefore, irrespective of shape and crystal structure, the edge, extracted from the integrated and averaged data, within the energy resolution range of ${\text{1\ eV}}$, is at ${\text{11.919\ keV}}$.
Thus, Au microcrystallites are in pure metallic form\cite{yang2005activation}.\\
Since the phase and amplitude are related by KKR, accurately measuring either contrast is generally sufficient for homogeneous materials, as the other contrast can be derived via the KKR.
In this work, we demonstrate the successful generation of XANES spectra with both phase and amplitude contrasts.
This eliminates the need for \textit{a priori} knowledge of the sample, making the spectro-ptychography broadly applicable.
\subsection*{Single-pixel high-resolution XANES}
To spatially resolve the absorption edge, the individual pixelated maps of ${\textit{fcc}}$ THH are considered (see \mbox{Figure \ref{fig5}}).
With increasing energy toward the absorption edge, the effect of absorption increases, so does the fluorescence intensity (\mbox{Figure \ref{fig5}a}), and transmission intensity (\mbox{Figure \ref{fig5}b,c}) is reduced.
Although both transmission maps from STXM and ptychographic reconstruction in \mbox{Figure \ref{fig5}b,c} show similar features, the latter possesses better resolution.
Phase maps across the absorption edge are also shown in \mbox{Figure \ref{fig5}d}.
The ${\textit{fcc}}$ THH has a non-uniform thickness (see \mbox{Figure \ref{fig3}b,c}) and therefore, at any energy, there is a non-homogeneous phase shift across the crystallite length.
Hence for better visualization, the phase map at 11.842 keV (\textit{i.e.}, the lowest scanned energy) is subtracted from the phase maps over the energy stack (see \mbox{Figure \ref{fig5}e}).
Evidently, the center of the THH crystallite (in \mbox{Figure \ref{fig5}e}) lightens up at ${\text{11.919\ keV}}$, illustrating the edge.
This allows for the localization of small changes across the full crystallite body and demonstrates the quantitativeness of X-ray ptychography in amplitude and phase contrasts.
Furthermore, single-pixel XANES spectra using various contrasts, as shown in \mbox{Figure \ref{fig5}f,g}, are extracted from the center of the THH (marked as a white square in \mbox{Figure \ref{fig5}d}).
In the single-pixel XANES, distinct spectral features are present in all contrasts and importantly, comparable to the integrated spectra as shown in \mbox{Figure S5a,b}.
The extracted absorption edge is at \text{11.919 keV} (see \mbox{Figure S6}), akin to the integrated spectra in \mbox{Figure S5a,b}, with a relative change in phase shift of ${\text{0.146\ rad}}$ at the peak maximum, \text{11.919 keV}.
$f_1$ derived from the reconstructed transmission spectrum is also overlayed in \mbox{Figure \ref{fig5}g} where the spectral features signify similarity with the phase shift spectrum along with the absorption edge at $\text{11.919 keV}$.
The $f_1$ derived from fluorescence and transmission (from both STXM and ptychographic reconstruction) contrasts are plotted along with the phase shifts (\mbox{Figure S6}) and the nicely overlayed spectra depict similarity for the $\sim$ 1.32\um{} thick crystallite.
Therefore, successful demonstrations of single-pixel XANES are shown from phase\slash transmission (ptychography) and fluorescence\slash transmission (STXM) contrasts with pixel sizes of $\sim$17\ and 40\,nm, respectively.\\
Similarly, single-pixel XANES spectra are extracted from small-sized thin crystallites in \mbox{Figure \ref{fig6}}a,b.
For the ${\textit{bc(o,t)}}$ $\text{THH}_2$, generated single-pixel XANES spectral information reveals the absorption edge is at ${\text{11.919\ keV}}$ (\mbox{Figure \ref{fig6}c,d}).
Besides, the XANES spectra from the bipyramid are well-resolvable and similar to $\text{THH}_2$ (see \mbox{Figure \ref{fig6}e,f}).
The changes in phase shift in both cases are $\text{0.048 and 0.043 rad}$, respectively.
Hence with the current measurement, single-pixel XANES spectra are successfully extracted from all the contrasts, including the thinner crystallites (also see \mbox{Figure. S7-S9}).
Importantly, the anomalous phase shifts for the studied crystallites are $\sim$\,9.6--11.1\%, even at the single pixel level.
${\textit{fcc}}$ and ${\textit{bc(o,t)}}$ crystallites exhibit similar absorption edges, reflecting electronic structures akin to that of bulk \textit{fcc} Au (see Figure S10).
Although several parameters may influence the catalytic activity and stability, the electronic structures play a crucial role and therefore, their detailed study with high-resolution is indispensable to pave the way for better design of future catalysts.\\
XANES spectra with both phase and amplitude contrasts remain detectable even in thinner regions of the crystallite (${\sim}\text{150 nm}$ thick, see Figure S11).
Notably, the absorption edge is still clearly resolved in the 60 nm thick bipyramid tips using amplitude contrast, despite the overall absorption being only ${\text{1.7}}{\%}$ (see Figure \ref{fig6}g).
In contrast, phase contrast spectrum in this region is noisy and unsuitable for accurately identifying the absorption edge.
These observations indicate that amplitude contrast offers greater sensitivity for X-ray absorption spectroscopy in the hard X-ray regime, even though phase contrast generally provides superior performance for imaging applications (see Figure \ref{fig2}d,e).
Owing to the non-homogeneous background present in the investigated sample and the inherent global phase ambiguity associated with ptychographic reconstruction, background contributions are subtracted from energy-dependent phase-contrast maps.
Nevertheless, residual background effects may persist, limiting sensitivity to small variations in phase-shift values.
Overall, the spectro-ptychography can be applied to diverse sample systems, including heterogeneous, multi-elemental materials and alloys.
Finally, Au microcrystallites hosting different morphologies, thicknesses and crystal structures, served as an ideal sample system for the demonstration of the capabilities and limitations of the spectro-ptychography.
Specifically, we learnt the \textit{bc(o,t)} THH crystallites are thicker than the \textit{bc(o,t)} bipyramid while there are no noticeable changes in the electronic structures of \textit{bc(o,t)} Au compared to bulk \textit{fcc} Au.
As a method, high-spatial resolution XANES with phase and amplitude contrasts are showcased in thin Au crystallites along with the feasibility of integrating it with other structural and chemical characterization methods.
While even minute changes can be captured using this technique, the present study reveals the electronic nature of the Au crystallites to be similar to that of bulk Au.
For a more comprehensive analysis of the sub eV range, a monochromator with smaller bandwidth of \makebox{sub-eV} would be required along with imaging capabilities.
With the advancement of synchrotron sources, sub-eV energy bandwidth with a small focus should be achievable in the upcoming years.
\section*{Conclusion}
In summary, we have successfully collected XANES spectra from individual Au microcrystallites using multimodal spectro-microscopy, including ptychographic phase, ptychographic amplitude, transmission, and fluorescence contrasts.
The $\text{Au} \, \mathrm{L}_3$ absorption edge is consistently identified across a wide range of crystallite sizes (100 nm to 2.4\um), morphologies and thicknesses ($\text{60 nm}$ to 1.71\um), demonstrating the robustness of the methods.
Single-pixel XANES enabled spatially resolved mapping of the absorption edge, achieving spatial resolutions of {$\sim$45\, nm} for ptychography and {$\sim$100\, nm} for XRF and STXM, with respective pixel sizes of 17 and \text{40 nm}.
Crystallites hosting both conventional and unconventional lattices exhibited a consistent absorption edge at \text{11.919 keV}, suggesting minimal influence of lattice type on the local electronic structures within $\text{1 eV}$ energy bandwidth.
It is intriguing that in spite of Au being in a strained foreign lattice, its local electronic structure is hardly affected.\\
This study establishes the feasibility of both amplitude and phase contrast XANES for ${\sim}\text{150 nm}$ thin Au crystallite in the hard X-ray energies, where X-ray absorption is typically weak. While phase-contrast imaging offers greater sensitivity for small and thin crystallites at the studied energies, amplitude contrast provides higher sensitivity to quantitative spectroscopic changes across the absorption edge.
Specifically, amplitude contrast can detect small changes even in a \text{60 nm} crystallite, despite a pre-edge attenuation of less than ${\text{2}}{\%}$ at \text{11.842 keV}.
Among the X-ray imaging modalities, ptychographic phase, ptychographic amplitude, and fluorescence contrasts are most sensitive to the $\text{Au} \, \mathrm{L}_3$ absorption edge across all crystallites.
In contrast, STXM-based transmission contrast is more effective in thicker crystallites, though ptychographic amplitude contrast offers the best combination of spatial resolution and spectral detail overall.
These results demonstrate the strength of phase and amplitude contrast XANES in the hard \linebreak X-ray regime and highlight spectro-ptychography as a superior technique among current X-ray spectro-microscopy methods.
This approach is especially well-suited for high-resolution investigations of catalysts and battery materials, including those based on Au, Ir and Pt, under \textit{ex-situ}, \textit{in-situ}, and \textit{operando} conditions.

\section*{Experimental Section}

\begin{figure}
	\centering
	\includegraphics[width=0.65\textwidth]{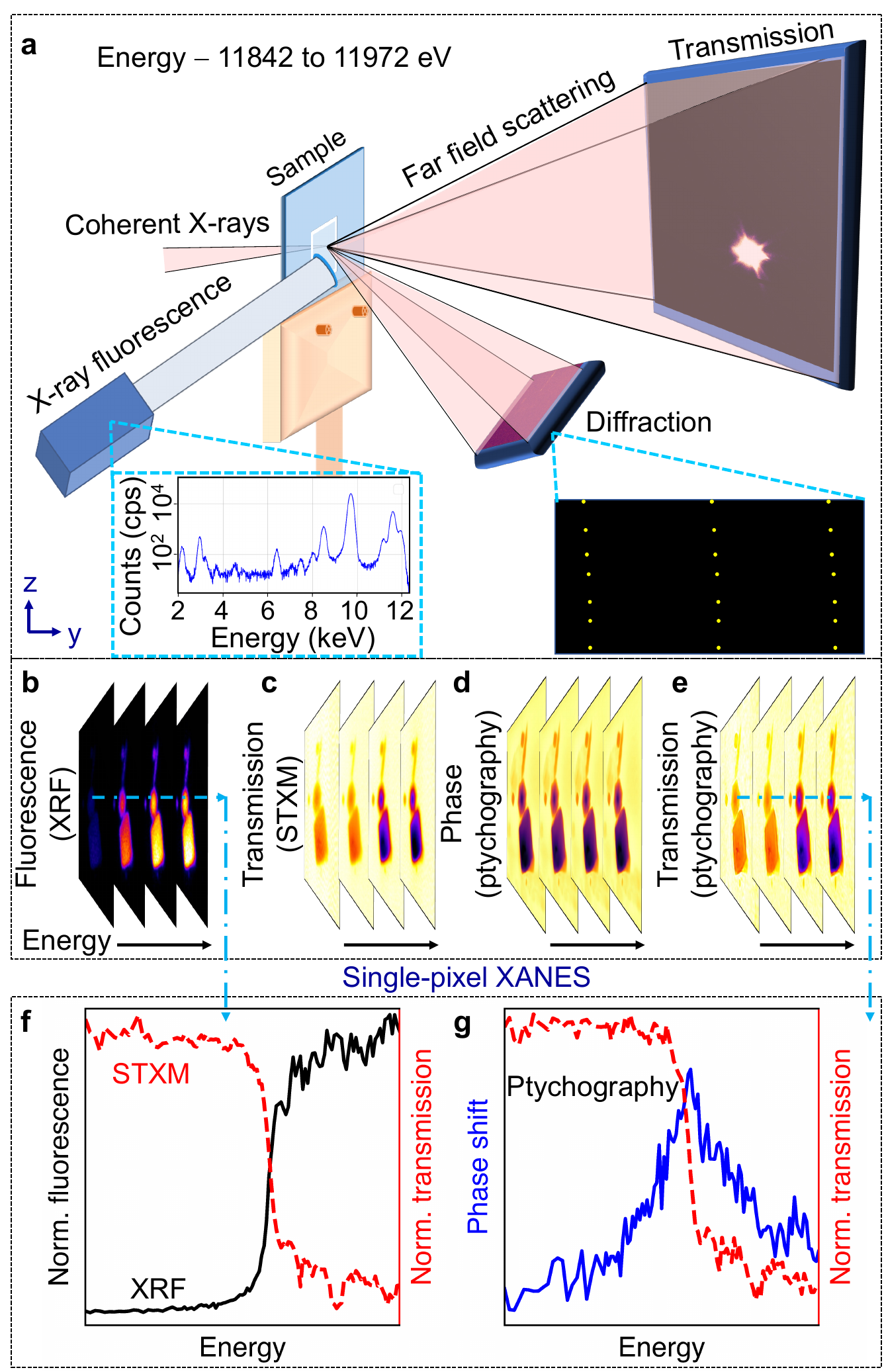}
	\caption{\textbf{Collection of X-ray absorption near edge spectroscopy data.}
		(a) Schematic representation of the experimental setup.
		A $\sim$100 nm beam was scanned across the sample in a 2D overlapping grid while simultaneously acquiring XRF, STXM and SXDM data.
		The energy of the beam was tuned from 11.842 to 11.972 keV spanning the Au $\mathrm{L}_3$ edge.
		At those energies, a series of (b) fluorescence, (c) transmission (STXM), ptychographically (d) reconstructed phase, and (e) reconstructed transmission maps are obtained.
		Illustrative single-pixel XANES spectra extracted from various contrasts: (f) fluorescence\slash transmission (STXM) and (g) ptychographic reconstructed phase\slash transmission.
		The wide-angle diffraction (SXDM) data is not presented.}
	\label{fig1} 
\end{figure}

\begin{figure}
	\centering
	\includegraphics[width=0.9\textwidth]{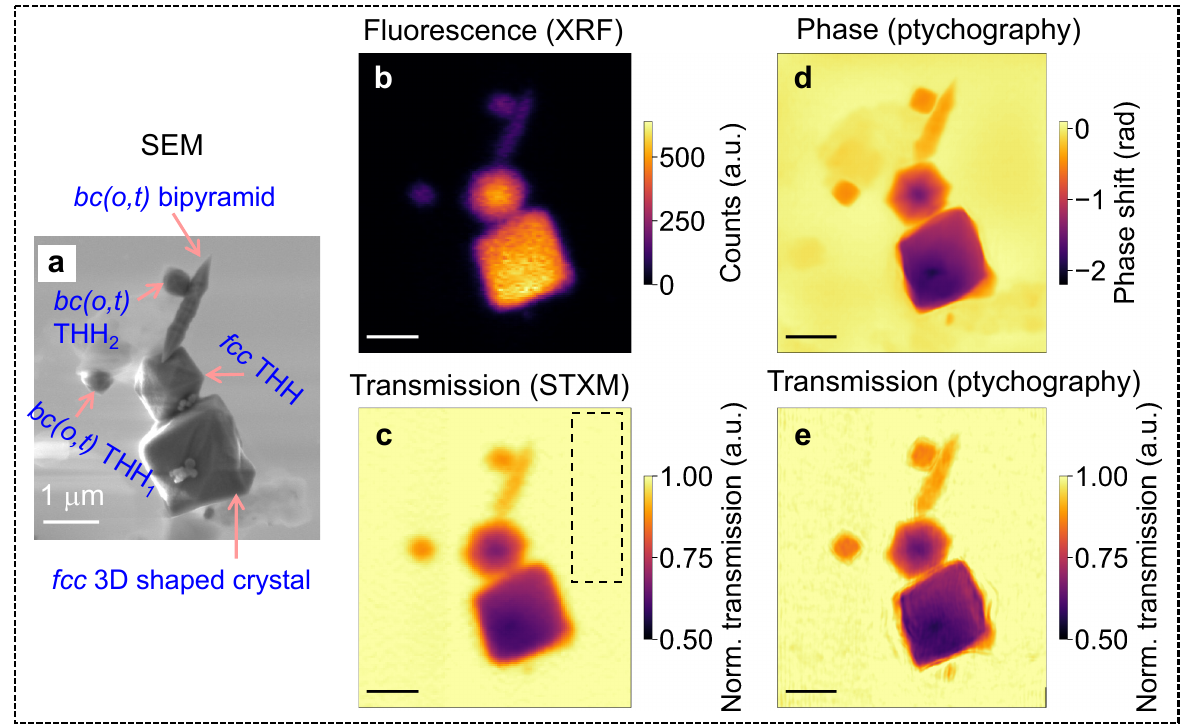}
	\caption{\textbf{Au microcrystallites under different contrasts.}
		(a) SEM image of a distribution of Au microcrystallites.
		(b-e) Scanning X-ray microscopy images with varying X-ray optical contrasts of the Au crystallites acquired at \text{E =11.963 keV}.
		(b) Fluorescence, (c) transmission, (d) reconstructed phase and (e) reconstructed transmission maps of the crystallites.
		Area within the black dashed box in (c) used for normalization of transmission maps.
		For reference, each crystallite has been defined by a unique name which is denoted in (a).
		The same nomenclature has been used throughout.
		Pixel sizes are \text{40 × 40\nm\textsuperscript{2}} (b,c) and \text{17 × 17\nm\textsuperscript{2}} (d,e).
		Scale bar, 1\um.
		THH refers to tetrahexahedral shaped crystallite. }
	\label{fig2} 
\end{figure}

\begin{figure}
	\centering
	\includegraphics[width=0.6\textwidth]{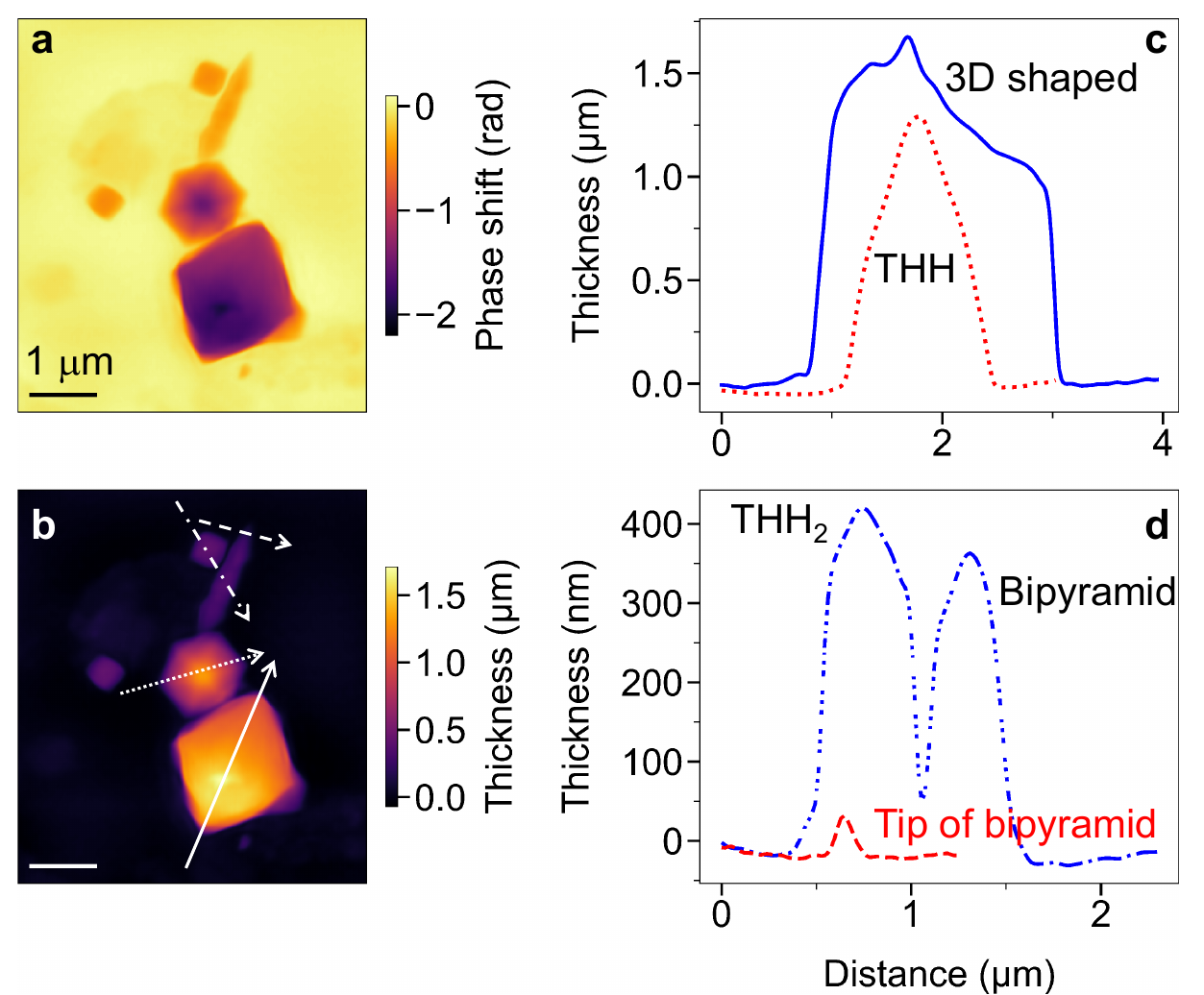}
	\caption{\textbf{Thickness calculation.}
		(a) Phase map of the crystallites at \text{E = 11.842} keV and (b) the corresponding calculated thickness map.
		The line profiles drawn over the crystallites in (b) are shown in (c,d).
		Distinctly, a higher phase shift relates to thicker crystallite.
		Scale bar, 1\um.}
	\label{fig3} 
\end{figure}

\begin{figure}
	\centering
	\includegraphics[width=0.9\textwidth]{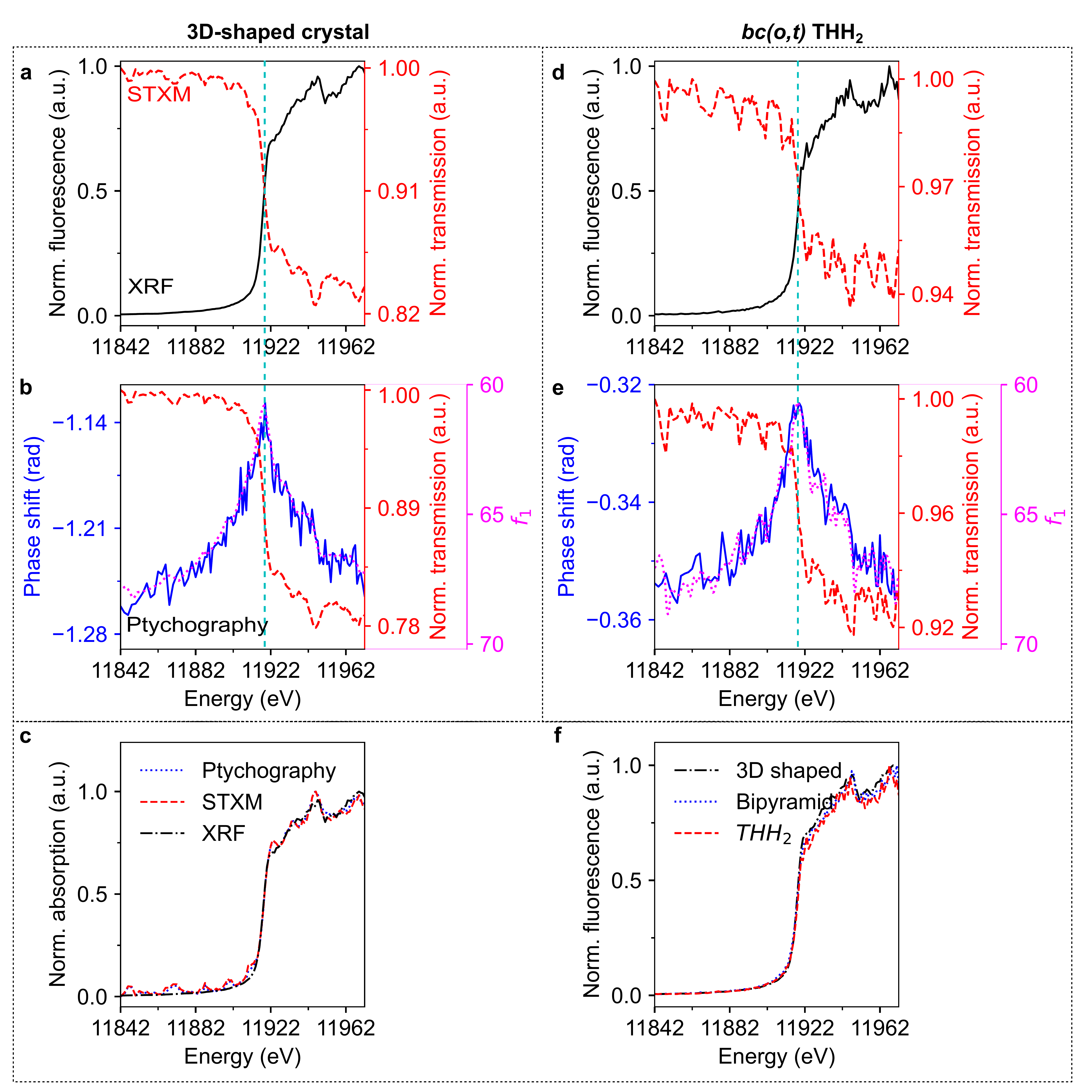}
	\caption{\textbf{Integrated and averaged XANES from various contrasts.}
		XANES spectra from (a,d) fluorescence\slash transmission (STXM) and (b,e) phase shift\slash transmission (ptychography) contrasts for ${\textit{fcc}}$ 3D shaped and ${\textit{bc(o,t)}}$ ${\text{THH}_2}$ crystallites, respectively.
		Calculated $f_1$ for (b) 3D-shaped and (e) $\text{THH}_2$ crystallites are also plotted.
		$f_1$ is derived using Kramers-Kronig transformation \cite{Watts2014Calculation} of transmission spectrum (ptychography) in (b,e).
		Cyan dotted line representing the edge is drawn for reference.
		(c) Comparison of XANES spectra collected using fluorescence and transmission (both, STXM and ptychography) contrasts from 3D-shaped crystallite. 
		(f) Comparison of fluorescence contrast XANES spectra for differently shaped crystallites.
		The integrated spectral information is extracted from the total area of each crystallite.
		Crystallite details are written at the top of the panel.
		}
	\label{fig4} 
\end{figure}

\begin{figure}
    \centering
    \includegraphics[width=0.9\textwidth]{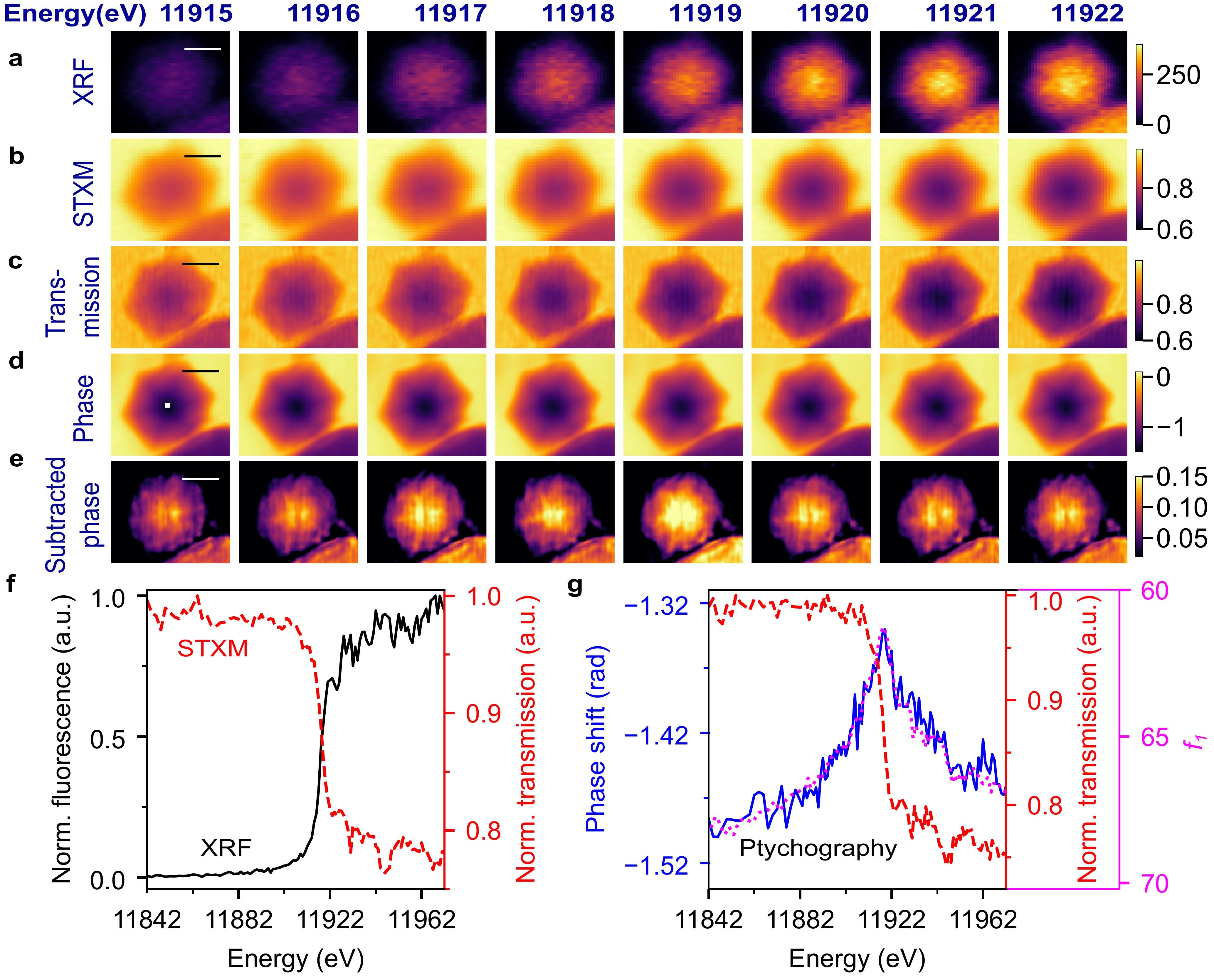}
    \caption{\textbf{Single-pixel XANES of ${\textit{fcc}}$ THH.}
		Energy-dependent (a) fluorescence, (b) transmission (STXM), (c) transmission (ptychography) and (d) phase (ptychography) maps of ${\textit{fcc}}$ THH crystallite across the $\mathrm{L}_3$ edge.
		(e) Subtracted phase map of the same crystallite.
		Non-homogeneous thickness across the crystallite generates a non-homogeneous phase shift.
		Therefore, the phase map at 11.842 keV has been subtracted from all the phase maps at different energies for better visualization.
		The center of the crystallite in (E) lightens up at 11.919 keV denoting the edge.
		Scale bar, 500 nm.
		(f,g) Single-pixel XANES spectra using fluorescence, transmission (STXM), phase and transmission contrasts, collected at the center, shown in a white box in (D).
		Derived $f_1$ from reconstructed transmission spectrum (ptychography) in (G) using KKR is also plotted in (G).
		Color bars: (A) counts (a.u.), (B,C) norm. transmission (a.u.) and (D,E) phase shift (rad).
		Pixel sizes are \text{40 × 40\nm\textsuperscript{2}} (A,B) and \text{17 × 17\nm\textsuperscript{2}} (C-E).
		Energy details are written at the top of the panel. }
	\label{fig5}
\end{figure}

\begin{figure}
	\centering
	\includegraphics[width=0.75\textwidth]{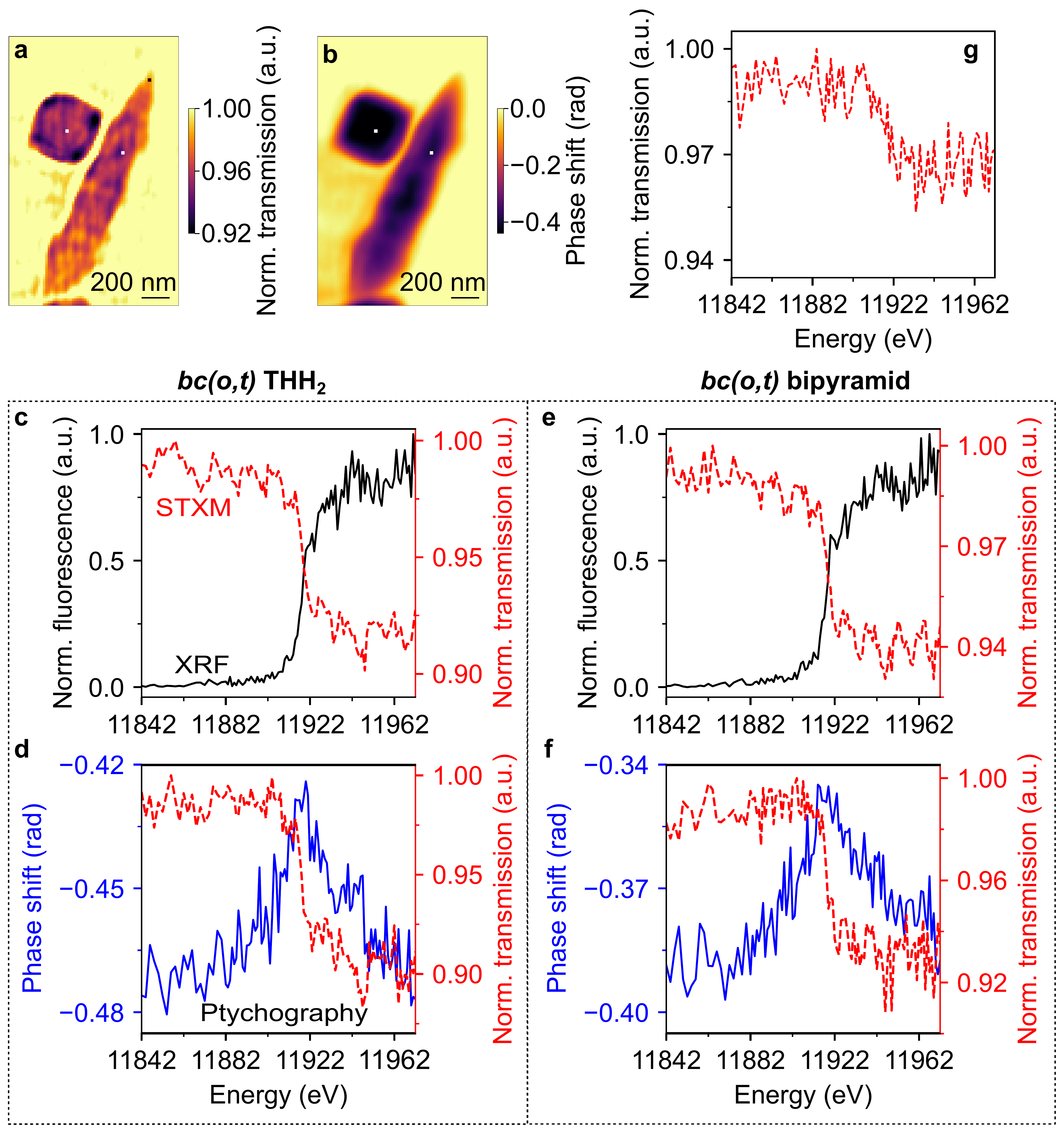}
	\caption{\textbf{Single-pixel XANES of ${\textit{bc(o,t)}}$ $\text{THH}_2$ and bipyramid.}
		(a,b) Ptychographic transmission and phase maps of the crystallites at \text{11.842 keV}. Single-pixel XANES spectra using (c,e) fluorescence\slash transmission (STXM) and (d,f) phase\slash transmission (ptychography) contrasts collected from $\text{THH}_2$ and bipyramid, respectively.
		The spectral information is extracted from the white boxes in (a,b).
		Pixel sizes are \text{40 × 40\nm\textsuperscript{2}} in XRF and STXM, and \text{17 × 17\nm\textsuperscript{2}} in ptychography.
		(g) Ptychographic transmission-contrast XANES extracted from the black squared region in bipyramid tip (a).
		}
	\label{fig6} 
\end{figure}

\textbf{Chemicals.} For the synthesis of the crystallites, tetraoctylammonium bromide (ToABr), silver nitrate ($\mathrm{AgNO_3}$) and hydrogen tetrachloroaurate(III) hydrate ($\mathrm{HAuCl_4}\!\cdot\!\mathrm{3H_2O}$) were obtained from Spectrochem, India.
Toluene was procured from S D-Fine Chem, India.
Double distilled water was used for the syntheses.
All solvents and reagents were used as procured without any further purification.
\\\textbf{Synthesis of Au microcrystallites.} The Au microcrystallites were synthesized following previous reports\cite{Mettela2013Highly,Sow2021Crystal}. 
Briefly, thermal decomposition of the AuAgToABr precursor followed by washing with acetone and aq. ammonia results in Au microcrystallites.
By altering the thermolysis temperature to $130\,^{\circ}\mathrm{C}$
results in \textit{fcc}\cite{Mettela2013Highly} while $250\,^{\circ}\mathrm{C}$ in \textit{bc(o,t)} crystallites\cite{Sow2021Crystal}.
\\\textbf{Sample preparation.} The as-prepared crystallites were suspended in acetone and then drop casted onto the
\mbox{Si/SiN\textsubscript{x}} membrane.
The crystallites loaded membrane was imaged under SEM\cite{stierle2016desy}.
In order to locate the crystallites for the X-ray microscopy experiment, X-shaped (16 × 16\um\textsuperscript{2})\ Pt markers were fabricated using a dual beam FIB/SEM\cite{stierle2016desy}.
The markers were located by the 100x magnification of the optical microscope available at the beamline. 
Further, star-shaped small markers (5 × 5\um\textsuperscript{2})\ were fabricated closer to the crystallites (see \mbox{Figure S2}).\\
\textbf{Performing the multi-modal spectro-microscopy experiment.} The multimodal spectro-microscopy experiment was performed at the PtyNAMi setup\cite{Schropp2020PtyNAMi} of beamline P06 at PETRA III, Deutsches Elektronen-Synchrotron (DESY)\cite{schroer2017ptynami}.
A Si(111) monochromator was used to obtain the monochromatic beam; the beam was focused by Si nano-focusing lenses (NFLs) and the downstream scattering from the lenses was shielded by a pinhole.
The energy bandwidth of the beam is approximately \text{1 eV} and the beam was focused down to \text{100 nm} as measured at \text{11.912 keV}.
In order to collect the X-ray fluorescence (XRF) signals, an energy-dispersive Vortex EM detector was placed ${{\sim}\text{1\ cm}}$ away from the sample at ${\text{90\textdegree}}$ with respect to the sample (see \mbox{Figure \ref{fig1}a}).
Besides, an in-vacuum \text{Eiger X 4M} detector was placed at ${{\sim}\text{3.5\ m}}$ downstream from the sample, to record the far-field diffraction patterns.
Additionally, the contribution of Bragg-diffraction, if any, was recorded using an \text{Eiger X 500k} detector (the data are not discussed here).
The sample was positioned in the focus and scanned in the y-z direction with a step size of 80 × 40\nm\textsuperscript{2} and exposure time of \text{20 ms} per scanning point.
Beam characterization (size and shape) was conducted via ptychography with a resolution test chart (model ATN\slash XRESO-50HC by NTT-AT)\cite{schroer2010hard, schroer2017ptynami, Schropp2020PtyNAMi}.
\\\textbf{Data analysis.} Data were analyzed using scripts written in Python. 
\\\underline{Transmission from STXM}: The transmission maps from STXM were generated by integrating the signals of the far-field diffraction patterns recorded on the \text{Eiger X 4M} detector.
The images were aligned using Fiji StackReg plugin.
The maps were normalized with respect to the mean of a region of the scan without sample (see Figure \ref{fig2}c).
Then, the maps were aligned using rigid and B-spline transformations\cite{Bron2014Fast, Ntatsis2023itk}.
For the normalized spectra, the spectra were divided by their highest intensity values of the transmission maps to obtain the range of spectra as 0 to 1.
\\\underline{Fluorescence from XRF}: The fluorescence spectra were fitted using PyMCA\cite{Sole2007Multiplatform} and the contribution of the Au $\mathrm{L}_{\upalpha}$ line was taken into consideration.
Initially, the images were aligned using Fiji StackReg plugin.
Later, the alignment correction values obtained from the transmission maps in STXM were applied in the corresponding fluorescence maps.
The normalized spectra were divided by their highest intensity values of the fluorescence maps to obtain spectra in the range between 0 to 1.
\\\underline{Ptychography}: For ptychography, the far-field diffraction patterns recorded on the \text{Eiger X 4M} detector were considered, cropping a \text{256 × 256} region around the forward direction.
1000 iterations of the ePIE algorithm were performed on the approximately 16,500 collected diffraction patterns (per energy step).
The probe function was retrieved simultaneously with the transmission object.
\\\underline{Phase from ptychography}: The reconstructed phase maps were aligned using Fiji StackReg plugin.
Then the feature and background were manually segmented and the background was fitted.
Later, the fitted background was subtracted from the overall phase maps including features.
Then the phase maps were aligned using rigid and B-spline transformations\cite{Bron2014Fast, Ntatsis2023itk}.
\\\underline{Background correction of ptychographic images}: Ptychographic reconstructions have a global phase ambiguity and exhibit long-range background fluctuations that need to be removed before quantitative analysis.
In addition, the presence of a non-homogeneous background (see \ref{fig2}D) also needs to be taken into account.
Therefore, a binary mask was created to separate the crystallite from the empty substrate; the long-range fluctuations of the background were then fitted with 2D Legendre Polynomials of up to the fifth order.
The fit was optimized using a genetic algorithm.
The fitted background was then subtracted from the entire image, including crystallite regions (see Figure S3).
\\\underline{Transmission from ptychography}: The amplitude maps from ptychographic reconstruction were aligned using Fiji StackReg plugin.
Then the amplitude maps were normalized with respect to the mean of the air region.
Then, the maps were aligned using rigid and B-spline transformations\cite{Bron2014Fast, Ntatsis2023itk}.
Then the square of the amplitude, \textit{i.e.}, transmission maps were generated.
For the normalized spectra, the spectra were divided by their highest intensity values of the transmission maps to obtain the range of spectra as 0 to 1.
\\\underline{Generation of integrated and average spectra}: The features were manually segmented. The contribution from the segmented region was integrated and averaged over the number of pixels in the region.
\\\underline{Calculation of $f_1$}: $f_1$ was derived using Kramers-Kronig transformation of energy dependent fluorescence and transmission (both STXM and ptychography) spectra  by a piecewise Laurent polynomial method\cite{Watts2014Calculation}.

\noindent
\\\textbf{Acknowledgements} \par 
\noindent We acknowledge DESY (Hamburg, Germany), a member of the Helmholtz Association HGF, for the provision of experimental facilities. 
Parts of this research were carried out at \makebox{PETRA III} (P06 beamline) and DESY NanoLab, and we thank Satishkumar Kulkarni for assistance with the microscopy at the DESY NanoLab.
Financial support by the Department of Science \& Technology, Government of India (DST) provided within the framework of the India@DESY collaboration is gratefully acknowledged. 

\clearpage

\pagenumbering{arabic} 
\renewcommand{\thepage}{S\arabic{page}} 

\begin{center}

{\Large\bfseries Supporting Information\par}
\vspace{0.5em}

{\Large\bfseries High-Resolution X-ray Spectro-Microscopy of Au Microcrystallites\par}
\vspace{1em}

Chaitali Sow\textsuperscript{1*},
Martin Seyrich\textsuperscript{1},
Johannes Hagemann\textsuperscript{1},
Giridhar U. Kulkarni\textsuperscript{2},
Thomas F. Keller\textsuperscript{1,3},
Christian G. Schroer\textsuperscript{1,3,4},
Andreas Schropp\textsuperscript{1}

\vspace{1em}

{\small
\textsuperscript{1} Centre for X-ray and Nano Science CXNS, Deutsches Elektronen-Synchrotron DESY, Notkestr. 85, Hamburg 22607, Germany\par
\vspace{0.5em}
\textsuperscript{2} Chemistry and Physics Of Materials Unit (CPMU), Jawaharlal Nehru Centre For Advanced Scientific Research (JNCASR), Rachenahalli Lake Rd, Jakkur, Bengaluru, Karnataka 560064, India\par
\vspace{0.5em}
\textsuperscript{3} Center for Hybrid Nanostructures, Institute for Nanostructure and Solid-State Physics, Universität Hamburg, Luruper Chaussee 149, 22761 Hamburg, Germany\par
\vspace{0.5em}
\textsuperscript{4} Helmholtz Imaging, Deutsches Elektronen-Synchrotron DESY, Notkestr. 85, Hamburg 22607, Germany\par
}

\vspace{2em}

{\small
\textsuperscript{*} Email:
\href{mailto:chaitali.sow@desy.de}{chaitali.sow@desy.de}, 
\href{mailto:chaitali.sow@xray-lens.de}{chaitali.sow@xray-lens.de}
\par
}
\end{center}

\setcounter{figure}{0}
\renewcommand{\thefigure}{S\arabic{figure}}

\begin{figure}
	\centering
	\includegraphics[width=0.8\textwidth]{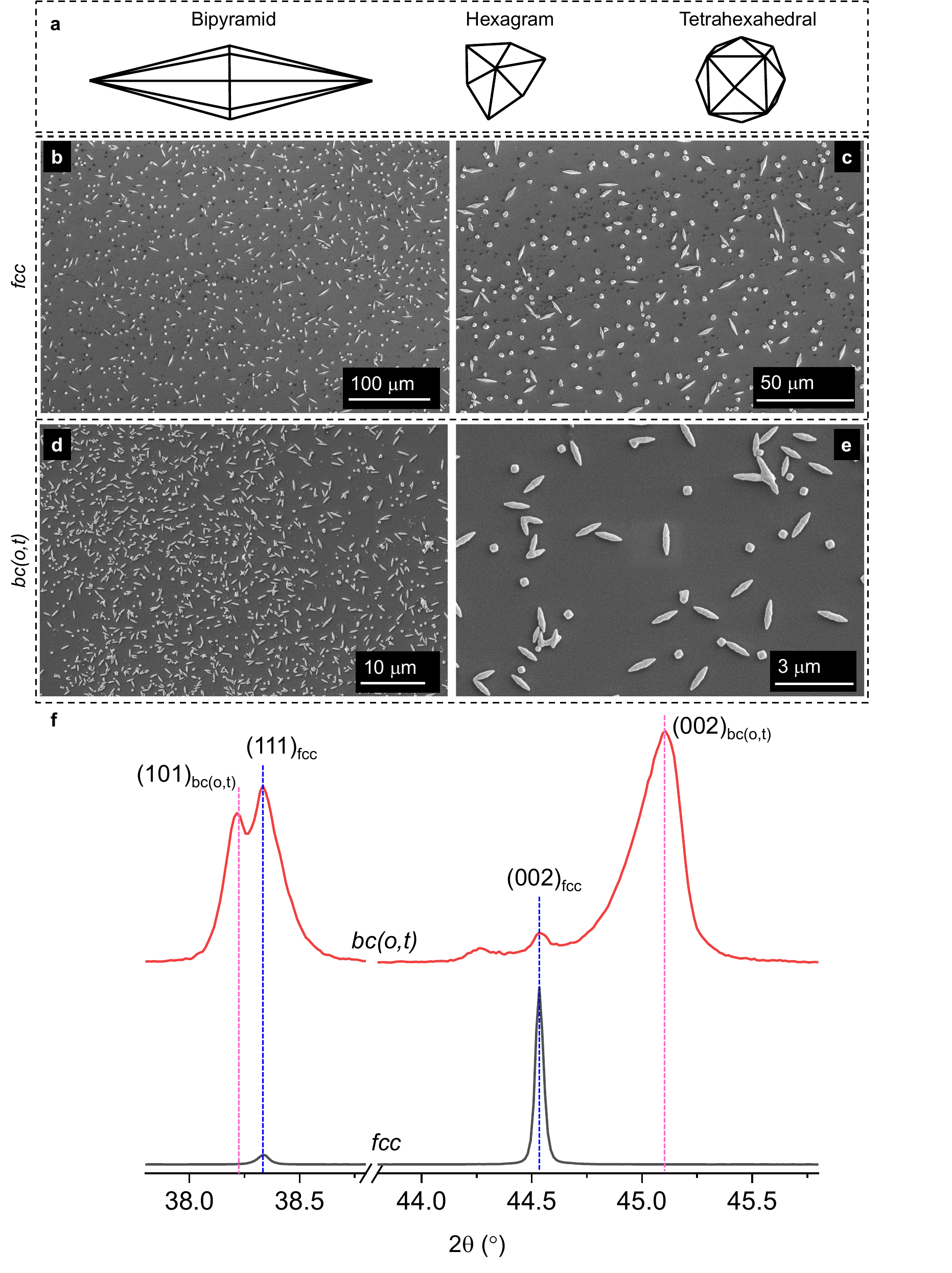}
	\caption{\textbf{\textit{fcc} and \textit{bc(o,t)} Au microcrystallites.}
		(a) Schematic representation of bipyramid, hexagram and tetrahexahedral shaped crystallites.
		SEM images of (b,c) ${\textit{fcc}}$ and (d,e) ${\textit{bc(o,t)}}$ Au microcrystallites at different magnifications.
		Note, the ${\textit{fcc}}$ crystallites are larger than the ${\textit{bc(o,t)}}$ crystallites.
		(f) XRD patterns of ${\textit{fcc}}$ and ${\textit{bc(o,t)}}$ crystallites.
		The patterns are normalized with respect to their maximum intensity peaks.}
	\label{figS1}
\end{figure}

\begin{figure}
	\centering
	\includegraphics[width=0.65\textwidth]{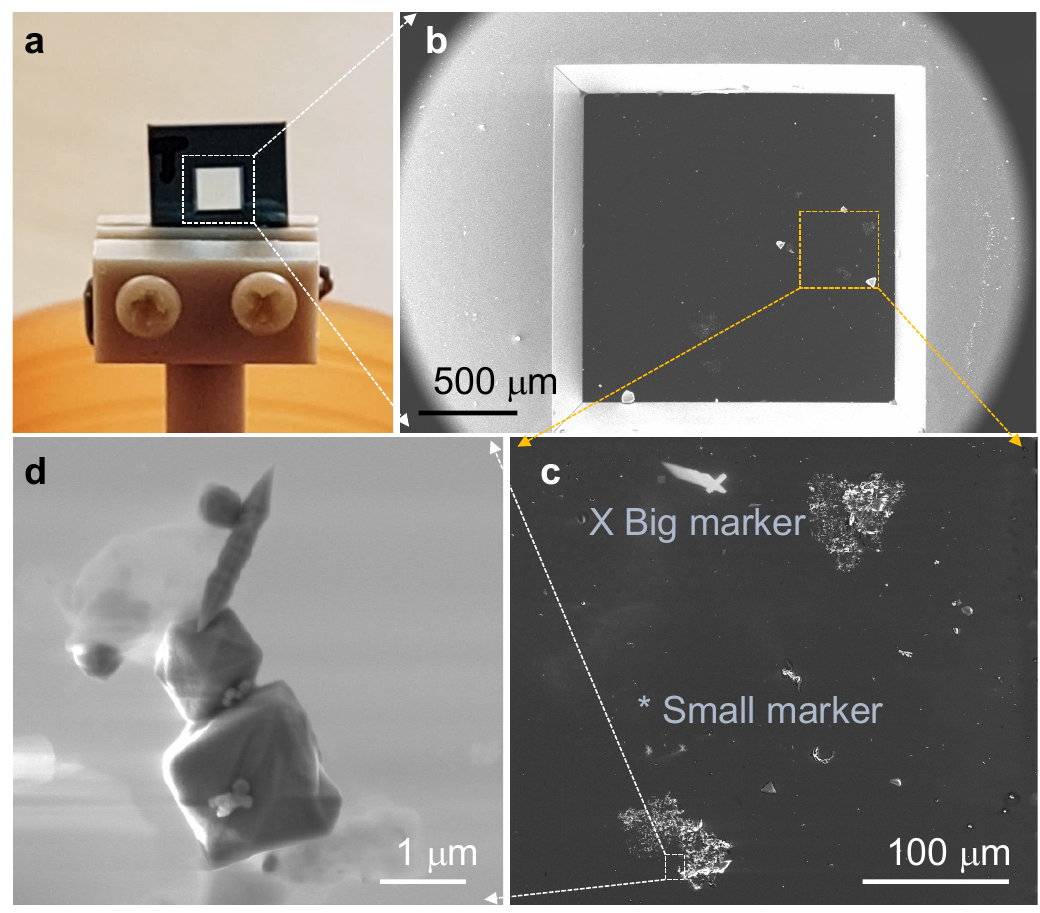}
	\caption{\textbf{Sample preparation for spectro-microscopy.}
		(a) Photograph of Si/SiN\textsubscript{x} membrane on the sample holder.
		(b-d) SEM images of the membrane, Pt markers, and Au microcrystallites under different magnifications.
		Different-sized Pt markers assisted in tracking the desired crystallites.}
	\label{figS2}
\end{figure}

\begin{figure}
	\centering
	\includegraphics[width=0.6\textwidth]{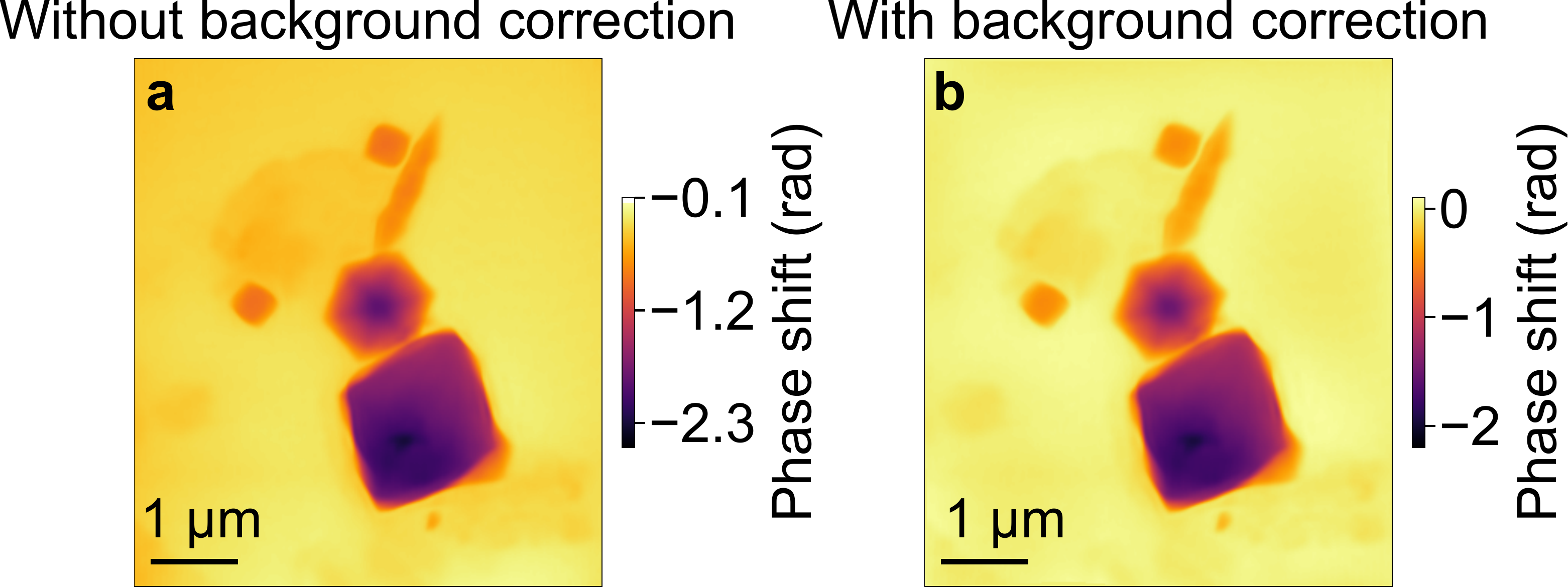}
	\caption{\textbf{Background correction of phase shift map.}
		Phase shift maps of Au crystallites at 11.842 keV, (a) without and (b) with background correction. }
	\label{figS3}
\end{figure}

\begin{figure}
	\centering
	\includegraphics[width=0.65\textwidth]{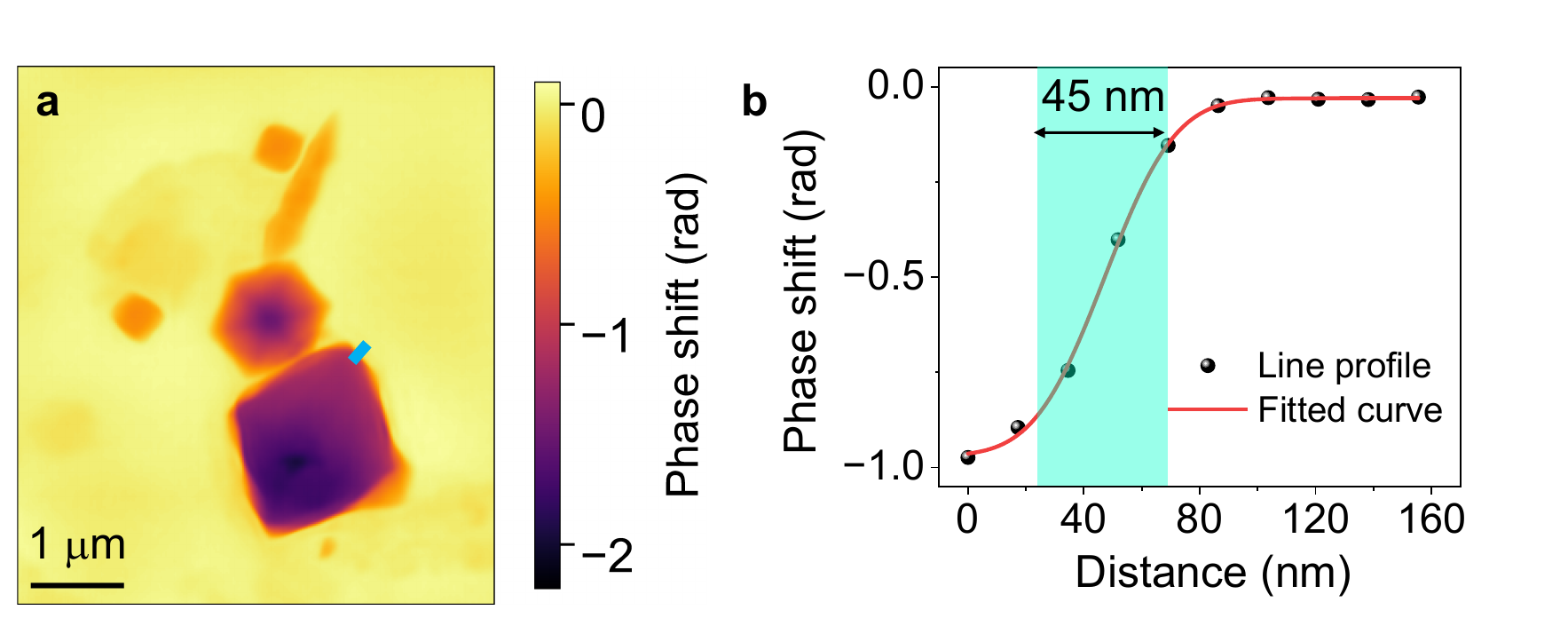}
	\caption{\textbf{Ptychographic resolution.}
		A line profile taken from the phase map (a) along the blue line is shown in (b). }
	\label{figS4} 
\end{figure}

\begin{figure}
	\centering
	\includegraphics[width=0.92\textwidth]{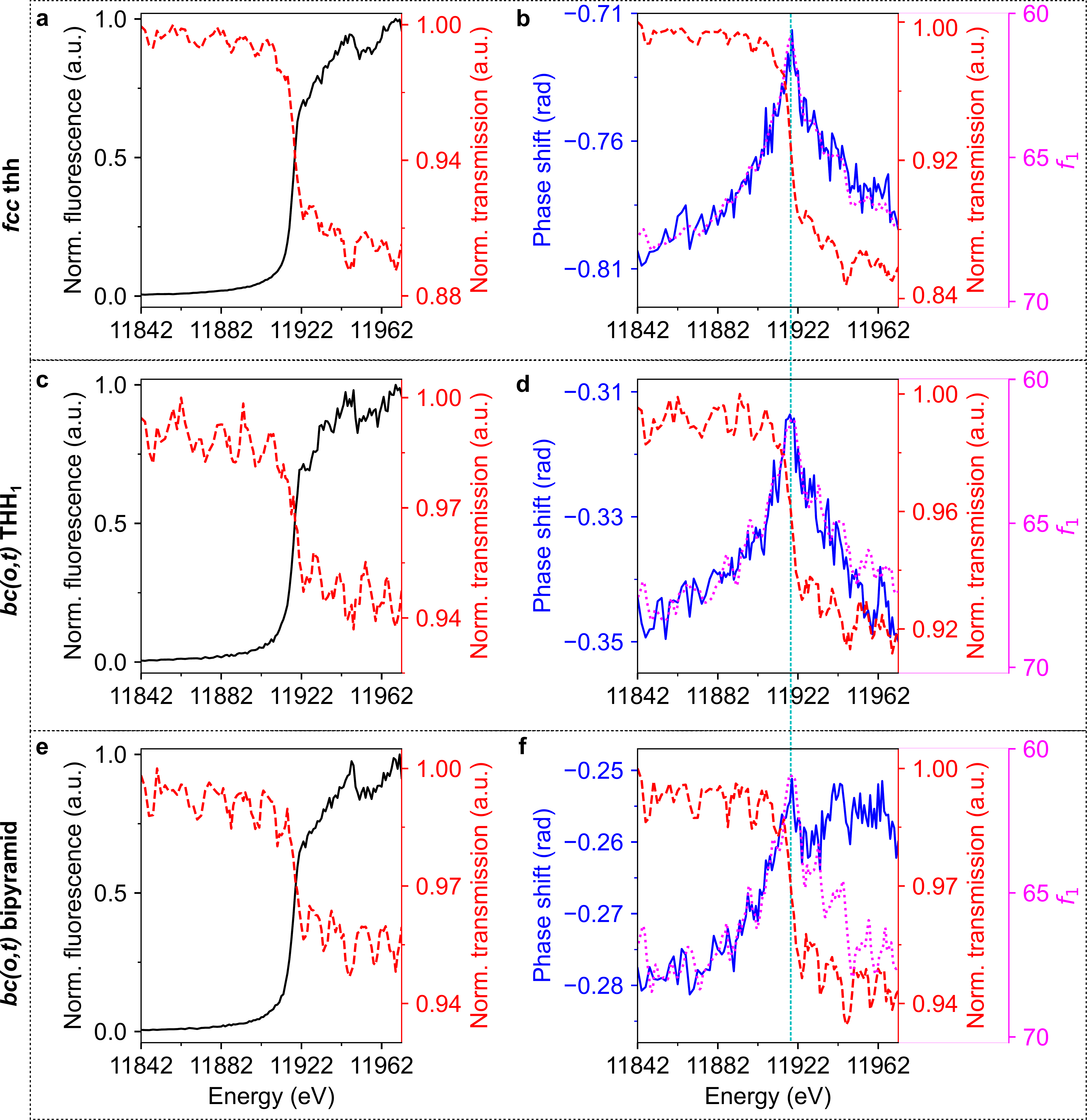}
	\caption{\textbf{Integrated and averaged XANES of THH, $\text{THH}_1$ and bipyramid.}
		Energy-dependent (a,c,e) fluorescence/transmission (STXM) and (b,d,f) phase shift, transmission (from ptychography) spectra and calculated $f_1$ for ${\textit{fcc}}$ THH, ${\textit{bc(o,t)}}$ $\text{THH}_1$ and ${\textit{bc(o,t)}}$ bipyramid, respectively.
		$f_1$ is derived from transmission spectra (ptychography) using Kramers-Kronig calculation. The peak positions are denoted by cyan dotted line.
		All spectral information is extracted from the integrated area and averaged over the number of pixels.
		Crystallite details are written at the left hand side of the panel.
		}
	\label{figS5}
\end{figure}

\begin{figure}
	\centering
	\includegraphics[width=0.5\textwidth]{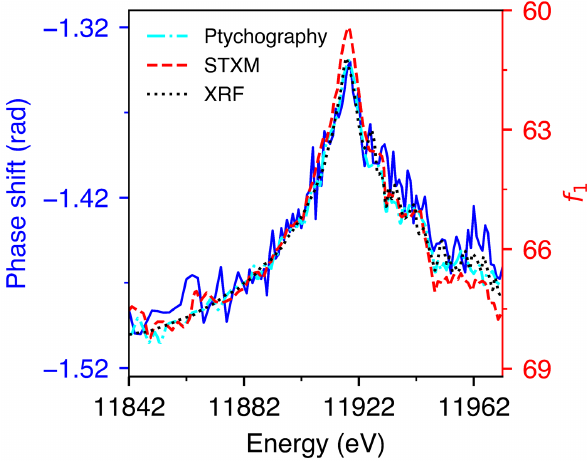} 
	\caption{\textbf{Identification of absorption edge in \textit{fcc} THH (from single pixel information).}
		Superimposed spectra of energy-dependent phase shift and calculated $f_1$.
		$f_1$ is generated from transmission spectra (ptychography and STXM) and fluorescence in Figure 5g,h.
		All spectral information is extracted from the single pixel value in Figure 5. }
	\label{figS6}
\end{figure}

\begin{figure}
	\centering
	\includegraphics[width=0.85\textwidth]{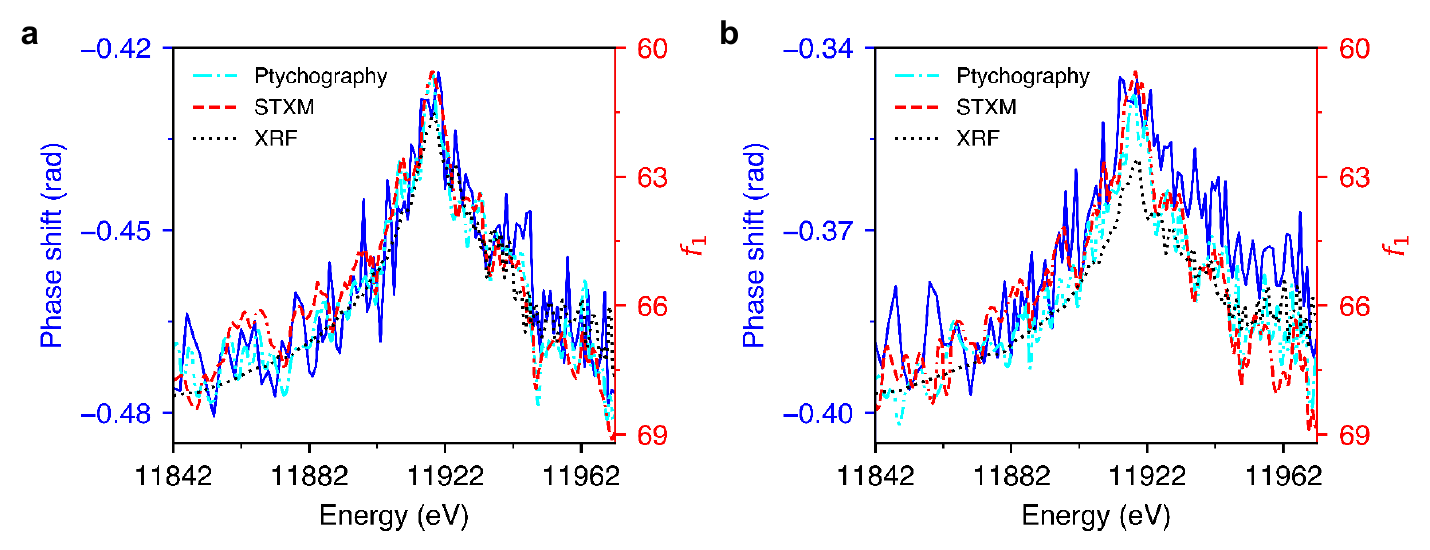}
	\caption{\textbf{Identification of absorption edge in $\text{THH}_2$ and bipyramid shaped crystallite (single pixel).}
		Identification of absorption edge in (a) $\text{THH}_2$ and (b) bipyramid shaped crystallites by superimposed phase shift and calculated $f_1$ spectra.
		$f_1$ is generated from transmission spectra (ptychography and STXM) and fluorescence in Figure 6 (single-pixel spectra).}
	\label{figS7}
\end{figure}

\begin{figure}
	\centering
	\includegraphics[width=0.92\textwidth]{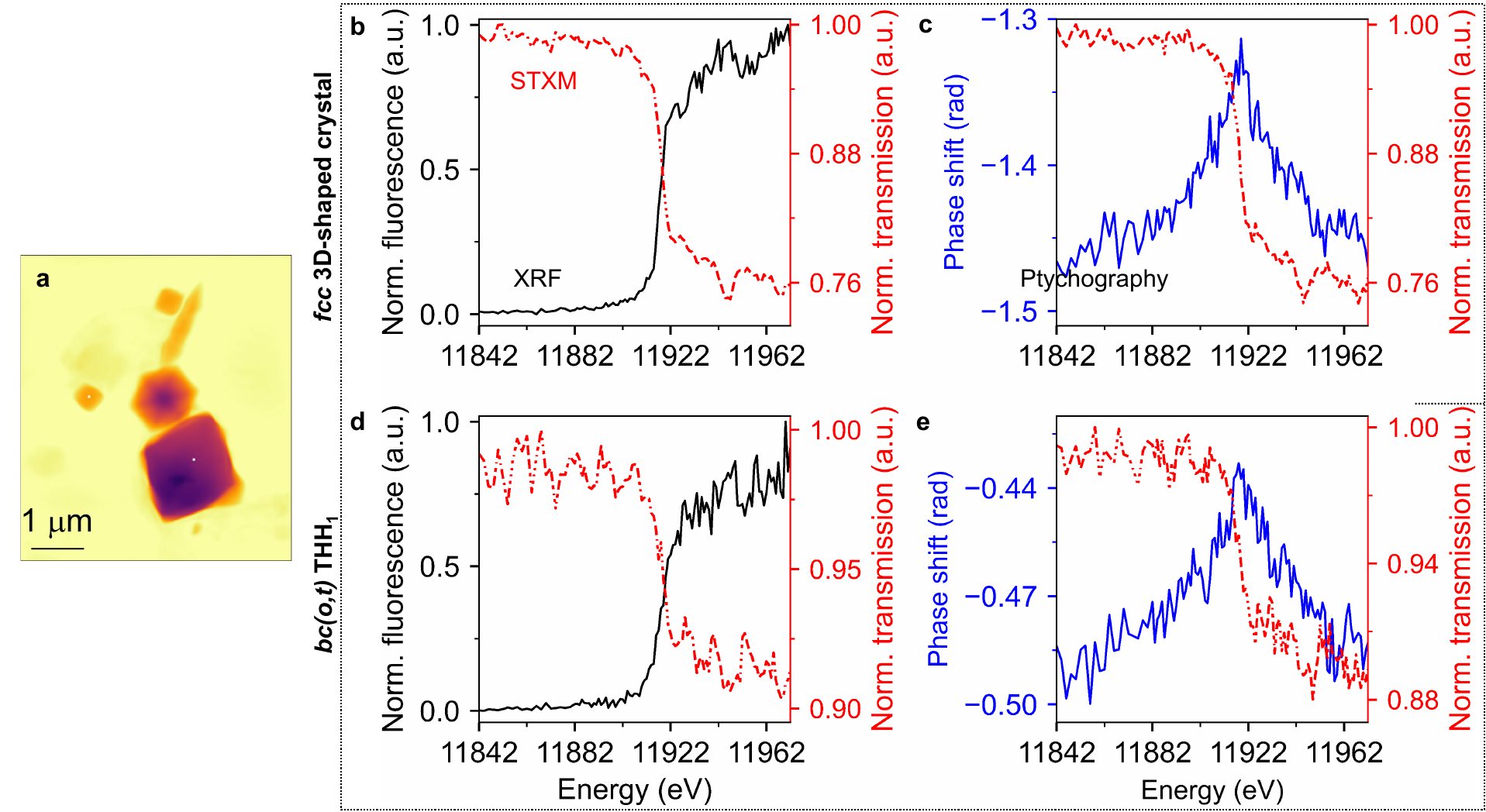} 
	\caption{\textbf{Single-pixel XANES of 3D-shaped crystal and $\text{THH}_1$.}
		(a) Phase map of the studied crystallites.
		Single-pixel XANES spectra using (b,d) fluorescence/transmission(STXM) and (c,e) phase/transmission (ptychography) contrasts collected from 3D-shaped crystal and $\text{THH}_1$, respectively.
		The spectral information are extracted from the white boxed regions in (a).
		Pixel sizes for (b,d) and (c,e) are \text{40 × 40\nm\textsuperscript{2}} and \text{17 × 17\nm\textsuperscript{2}}, respectively. }
	\label{figS8}
\end{figure}

\begin{figure}
	\centering
	\includegraphics[width=0.8\textwidth]{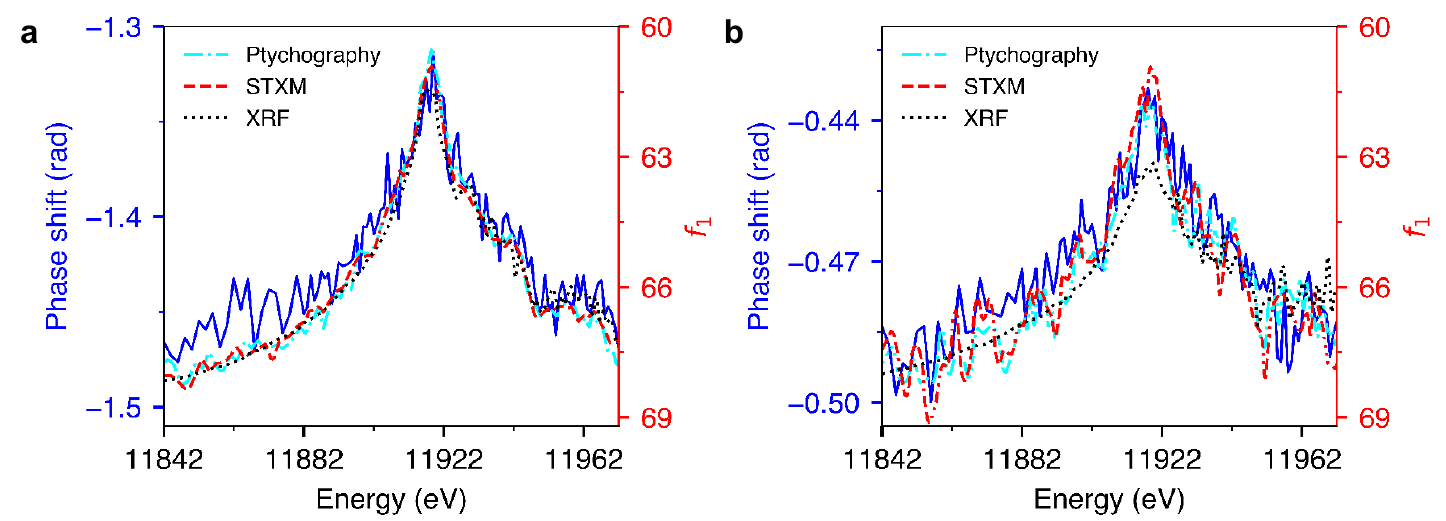}
	\caption{\textbf{Identification of absorption edge in 3D-shaped and $\text{THH}_1$ shaped crystallites (single pixel).}
		Identification of absorption edge in (a) 3D-shaped and (b) $\text{THH}_1$ shaped crystallites by superimposed phase shift and  calculated $f_1$ spectra.
		$f_1$ is generated from transmission spectra (ptychography and STXM) and fluorescence in Figure S8 (single-pixel spectra). }
	\label{figS9}
\end{figure}

\begin{figure}
	\centering
	\includegraphics[width=0.92\textwidth]{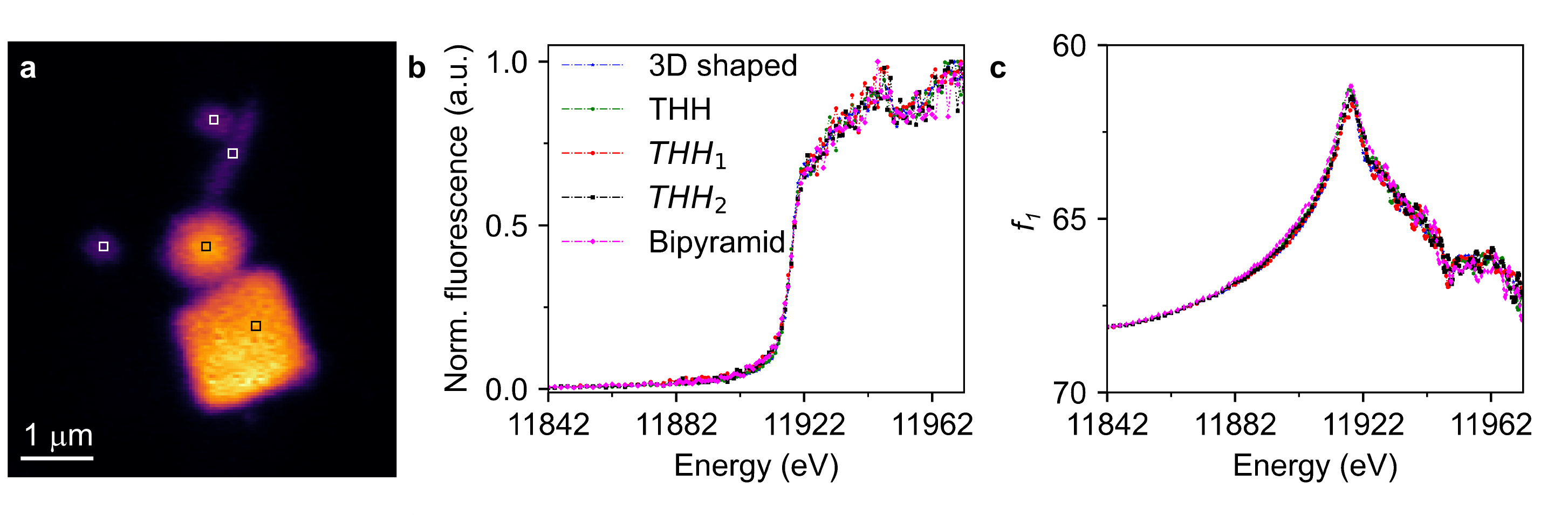}
	\caption{\textbf{Comparison of XANES of different shaped crystallites.}
		(a) XRF image of Au crystallites.
		(b) Norm. fluorescence spectra obtained from the central part of the crystallites with an area of \text{160 × 160\nm\textsuperscript{2}}, as marked in the XRF map in (a).
		(c) $f_1$ spectra derived from XANES spectra in (b).}
	\label{figS10}
\end{figure}

\begin{figure}
	\centering
	\includegraphics[width=0.7\textwidth]{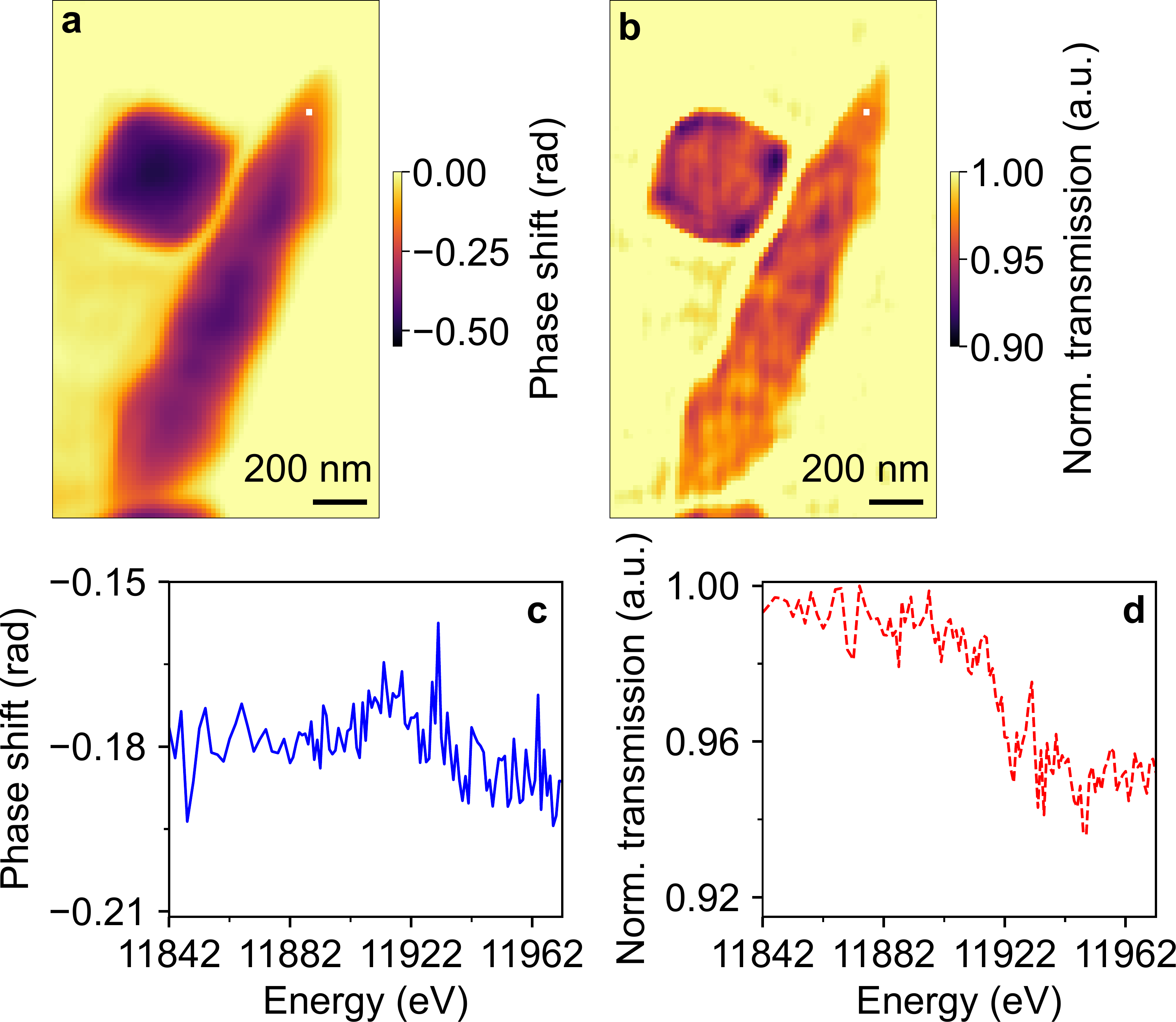}
	\caption{\textbf{Single-pixel XANES from approximately 150 nm thick bipyramid tip.}
		(c,d) Single-pixel XANES spectra extracted from the white boxed region in the (a) phase and (b) transmission (ptychography) maps, respectively.
		Pixel size: \text{17 × 17\nm\textsuperscript{2}}. }
	\label{figS11}
\end{figure}

\end{document}